\documentclass[prb, reprint, twocolumn, showpacs]{revtex4-1}

\usepackage{graphicx}
\usepackage{color}
\usepackage{amsmath}
\usepackage{amssymb}
\usepackage{amsthm}
\usepackage{stmaryrd}
\usepackage{hyperref}
\usepackage{soul}

\newcommand{\mathsym}[1]{{}}

\begin{document}

\title{Spin Hall and spin Nernst effects in a two-dimensional electron gas with Rashba spin-orbit interaction: temperature dependence}
\author{A. Dyrda\l$^{1}$,   J.~Barna\'s$^{1,2}$, and V. K.~Dugaev$^{3}$}

\address{$^1$Faculty of Physics, Adam Mickiewicz University,
ul. Umultowska 85, 61-614 Pozna\'n, Poland \\
$^2$  Institute of Molecular Physics, Polish Academy of Sciences,
ul. M. Smoluchowskiego 17, 60-179 Pozna\'n, Poland\\
$^3$Department of Physics and Medical Engineering, Rzesz\'ow University of Technology,
al. Powsta\'nc\'ow Warszawy 6, 35-959 Rzesz\'ow, Poland
}
\date{\today }

\begin{abstract}
Using the Matsubara Green function formalism we calculate the temperature dependence of spin Hall and spin Nernst conductivities of a two-dimensional electron gas with Rashba spin-orbit interaction in the linear response regime. In the case of spin Nernst effect we also include the contribution from spin-resolved orbital magnetization, which assures correct behavior of the spin Nernst conductivity in the zero-temperature limit.  Analytical formulas for the spin Hall and spin Nernst conductivities are derived in some specific situations. Using the Ioffe-Regel localization criterion, we have also estimated the range of parameters where the calculated results for the spin Hall and spin Nernst conductivities are applicable. Analytical results show that the vertex correction  totally suppresses the spin Hall conductivity at arbitrary temperature. The spin Nernst conductivity, in turn, vanishes at $T=0$  when the orbital contribution is taken into account, but generally is nonzero at finite temperatures.
\end{abstract}

\maketitle

%----------------------------------------
\section{Introduction}
%----------------------------------------

Spin Hall effect (SHE) was predicted theoretically in 1971 by Dyakonov and Perel. \cite{Dyakonov1971,Dyakonov1971_2}
Two decades later, interest in this effect was renewed after the paper by Hirsch,\cite{Hirsch1999} and now it attracts a lot of attention as a possible method of spin current generation.\cite{Murakami2003,SinovaUniversalSHE,EngelRashbaHalperin} The phenomenon was observed experimentally only a decade ago,\cite{Kato2004,Wunderlich2005} and since then it is one of the most important issues in spintronics, playing a major role in the process of generation and detection of spin currents in a pure electrical manner.\cite{sinova2012,sinova2014}
Spin current, in turn, plays a key role in the process of electrically controlled magnetic switching as it generates a spin torque when being absorbed in a magnetic system.\cite{LiuMoriyama2011,LiuLee2012,Pai2012}

It is also well known that the SHE is a consequence of spin-orbit interaction. This interaction, in turn, may be either of intrinsic or extrinsic origin. The extrinsic SHE is associated with mechanisms of spin-dependent electron scattering on impurities and other structural defects with spin-orbit interaction (skew scattering and side jump), whereas the intrinsic SHE is a consequence of a nontrivial trajectory of charge carriers in the momentum space due to the contribution of intrinsic spin-orbit interaction  to the corresponding band structure.\cite{sinova2014,Vignale2010,EngelRashbaHalperin}

The transverse spin currents and spin accumulation in systems with spin-orbit interaction can be induced  not only by an external electric field, but also due to a temperature gradient in the system. Thus, a difference in temperatures of the two ends of a system with spin orbit interaction gives rise to a spin current that flows perpendicularly to the temperature gradient. This phenomenon is known as the spin Nernst effect (SNE), but it is also sometimes referred to as the thermally-induced spin Hall effect.
\cite{Cheng2008,LiuXie,Ma}
Physical origin of SNE is very similar to that of SHE, and one can distinguish the intrinsic and extrinsic contributions to SNE, similarly as in the case of SHE. However, there are some differences between these two phenomena, which are associated with different influence of the driving forces (electric field for SHE and temperature gradient for SNE) on the electron distribution function in the momentum space. This difference leads to some differences in the relevant theoretical formalisms and descriptions.

As the physical mechanisms of the SHE and SNE are rather well understood, the influence of finite temperature on both these phenomena, especially on the  SHE, received less attention in the literature. The effect of temperature on the spin current was considered, for instance, by Bencheikh and Vignale,\cite{Vignale2008} who applied the spin density matrix method to a spin current in two-dimensional electron gas (2DEG) with Rashba and Dresselhaus spin-orbit interactions. Using a simple argument it was also predicted that the spin Hall conductivity in 2DEG with Rashba interaction should vanish at arbitrary temperature.\cite{Dimitrova2005,Erlingsson2005,Schliemann2006}
In turn, since a temperature gradient is the driving force for SNE, the temperature effects are inherently included in the  corresponding descriptions. Though many of them are limited to the low temperature regime, the zero temperature limit, however,  was not studied thoroughly enough.

There are several different approaches to the spin Nernst phenomenon. For example, Lyapilin~\cite{Lyapilin}  used the nonequilibrium statistical operator, taking into account electron scattering processes in 2DEG with Rashba interaction. This theory, however, does not give clear information on the role of individual microscopic mechanisms of SNE.
Similar problem was also studied by Ma.~\cite{Ma}  In turn,
Akera and Suzuura~\cite{Akera} used the Boltzmann equation formalism to analyze the SNE due to extrinsic mechanisms, like skew scattering and side jump. The spin Nernst effect due to skew-scattering was also considered by {\it{ab initio}} numerical methods based on various approximations with respect to the electronic structure and transport properties -- see for example Refs~[\onlinecite{Tauber2013,Wimmer2013,Zimmermann2014,Kovacik2015}].
The room temperature SNE was considered recently by T\"{o}lle {\it et al}.~\cite{Tolle2014} They have taken into account dynamical spin-orbit coupling (spin-orbit coupling with the vibrating lattice and impurities) and found a contribution due to the so-called dynamical side-jump mechanism. In turn, Borge {\it et al}\cite{Borge} used Matsubara Green functions to derive some results on thermoelectrics and spin thermoelectrics in two-dimensional disordered electron gas.
One should also mention, that the spin Nernst effect in the mesoscopic regime was studied using the Landauer-B\"uttiker formalism for the four-terminal cross-bar geometry in systems based on the 2DEG with Rashba spin-orbit coupling (see e.g. [\onlinecite{LiuXie}] and references therein) and HgTe quantum wells.~\cite{RotheHankiewicz}

There is still lack of a consistent theory that takes into account all microscopic mechanisms and describes the spin Hall and spin Nernst effects at high temperatures. This problem, however, is of significant importance if one takes into account experimental side and also possible applications, e.g. for magnetic switching in spintronics devices.
The importance of finite-temperatures in the theoretical interpretation of SHE has been  emphasized recently by Gorini {\it at al}.\cite{Gorini2015}
Moreover, the zero-temperature limit of the spin Nernst conductivity has not been studied in detail, though it is known that orbital contribution to the anomalous Nernst effect has to be included to have accurate zero temperature behavior.\cite{obraztsov,gusynin}
Therefore, in this paper we consider the temperature dependence of SHE and SNE. We restrict consideration to a two-dimensional electron gas with spin-orbit interaction of Rashba type. Such a system is a model system for many semiconductor devices. Using the Matsubara Green function formalism and the  auxiliary vector potential in the case of temperature gradient,~\cite{Dyrdal2013} we find some general equations that  describe the spin Hall and spin Nernst conductivities. In the latter case we also predict a contribution to the spin Nernst conductivity due to spin-resolved orbital magnetization.
Based on these equations, we have derived analytical formulas for the SHE and SNE conductivities in the {\it bare bubble} approximation. Taking into account electron scattering on impurities, we show that the total cancelation of the intrinsic contribution to the spin Hall conductivity  by the corresponding contribution due to vertex correction  takes place not only at zero temperature, but also at higher temperatures, in agreement with earlier predictions.\cite{Dimitrova2005,Erlingsson2005,Schliemann2006} In the case of spin Nernst effect we show that the total spin Nernst conductivity vanishes at $T=0$ when a spin-resolved orbital magnetization is taken into account, but generally is finite at nonzero temperatures.

This paper is organized as follows. In Section II we introduce the model of a two-dimensional Rashba electron gas and derive some general formulas for the spin Hall and spin Nernst conductivities. In Sections III and IV we present results on the spin Hall and spin Nernst  conductivities, respectively. The vertex corrections are also calculated there. In the case of spin Nernst effect we also included a contribution from spin-resolved orbital magnetization. Summary and final conclusions are given in Section V.

%----------------------------------------
\section{Model and method}
%----------------------------------------

Two-dimensional electron gas with parabolic dispersion relation is the simplest model used to describe electronic states in semiconductor heterostructures and quantum wells.  Rashba spin-orbit interaction in such systems appears due to the lack of structural inversion symmetry. Hamiltonian describing 2DEG with Rashba spin-orbit coupling, written in the plane-wave basis, takes the form
\begin{equation}
\label{1}
\hat{H} = \frac{\hbar^{2} k^{2}}{2 m} \sigma_{0} + \alpha (k_{y}\, \sigma_{x} - k_{x}\, \sigma_{y}),
\end{equation}
where  $\sigma_{n}$ for  $n = \{x, y, z\}$ are the Pauli matrices, while $\sigma_{0}$ is the unit matrix -- all defined in the spin space. The parameter $\alpha$ in the above equation describes strength of the Rashba interaction, while $k_x$ and $k_y$ are the in-plane wavevector components. Eigenvalues of the Hamiltonian (1) have the form,  $E_{\pm} = \varepsilon_{k} \pm \alpha k$, with $\varepsilon_{k} = \hbar^{2} k^{2}/2 m$ and $k^{2} = k_{x}^{2} + k_{y}^{2}$.

Below we derive some general expressions for the dynamical spin Hall and spin Nernst conductivities in terms of the appropriate Green functions.  These expressions will be used in the subsequent sections to calculate the static conductivities.

\subsection{Spin Hall effect}

Spin Hall effect is a phenomenon in which external electric field drives a spin current that flows perpendicularly to the driving field. In the model description we assume a time-dependent external electromagnetic field of frequency $\omega/\hbar$ (here $\omega$ is energy) described by the vector potential
$\mathbf{A}(t)=\mathbf{A}(\omega)\exp (-i\omega t/\hbar) $.
The corresponding electric field is then related to $\mathbf{A}$ {\it via} the formula $\mathbf{A}(\omega)=(\hbar /i\omega )\mathbf{E}(\omega)$.  The
perturbation part of the Hamiltonian due to interaction of the system with the external field, $H_{\mathbf{A}}^{\scriptstyle{E}}$, takes the form
\begin{equation}
\label{3}
\hat{H}_{\mathbf{A}}^{\scriptstyle{E}}(t) = - e \hat{\mathbf{v}}\cdot\mathbf{A}(t) \equiv - \hat{\mathbf{j}}^{el}\cdot\mathbf{A}(t),
\end{equation}
where $e$ is the electron charge ($e < 0$), $\hat{\mathbf{v}}$ is the electron velocity operator, $\hat{\mathbf{v}}=(1/\hbar )\partial \hat{H}/\partial \mathbf k $, while $\hat{\mathbf{j}}^{el} = e \hat{\mathbf v}$ is the operator of electric current density. The
$x$ and $y$ components of the velocity operator have the following explicit form:
\begin{eqnarray}
\label{7}
\hat{v}_{x} = \frac{\hbar k}{m} \cos(\phi) \sigma_{0} - \frac{\alpha}{\hbar} \sigma_{y},\\
\label{8}
\hat{v}_{y} = \frac{\hbar k}{m} \sin(\phi) \sigma_{0} + \frac{\alpha}{\hbar} \sigma_{x},
\end{eqnarray}
where $\phi$ is the angle between the wavevector $\bf k$ and the axis $x$, i.e. $k_x=k\cos (\phi)$ and $k_y=k\sin (\phi)$, while the last terms represent the components of the anomalous velocity.

Without loss of generality, we assume in this paper that the external electric field is oriented along the $y$-axis and calculate the spin current flowing along the $x$ axis.
The corresponding operator of spin current  density is defined as an anticommutator of the velocity operator and the $z$-th component of spin operator ($\hat{s}_{z} = \frac{\hbar}{2} \sigma_{z}$),
%: $\hat{j}_{x}^{s_{z}} = [\hat{v}_{x}, \hat{s}_{z}]_{+}/2$.
\begin{equation}
\hat{j}_{x}^{s_{z}} = \frac{1}{2} \left[\hat{v}_{x}, \hat{s}_{z} \right]_{+} =  \frac{\hbar^{2}}{2 m} k_{x} \sigma_{z}.
\end{equation}
The expectation value of the spin current induced by an external electric field can be found in the Matsubara-Green functions formalism  from the following formula:
\begin{equation}
\label{2}
j_{x}^{s_{z}} (i \omega_{m})= \frac{1}{\beta}  \sum_{\mathbf{k}, n} \mathrm{Tr}\left\{\hat{j}_{x}^{s_{z}} G_{\mathbf{k}}(i \varepsilon_{n} + i \omega_{m}) \hat{H}_{\mathbf{A}}^{\scriptstyle{E}} (i \omega_{m})G_{\mathbf{k}}(i \varepsilon_{n}) \right\},
\end{equation}
where $\beta =1/k_BT$, with $T$ and $k_B$ denoting the temperature and Boltzmann constant, respectively. The graphical diagram corresponding to Eq.(6) is presented in Fig.1a. The  perturbation term takes now the form $\hat{H}_{\mathbf{A}}^{\scriptstyle{E}}(i \omega_{m})=-e\hat{v}_yA_y(i \omega_{m})$, with the amplitude of the vector potential $A_{y}(i \omega_{m})$ determined by the amplitude $E_y(i \omega_{m})$ of electric field through the relation $A_{y}(i \omega_{m}) = \frac{E_{y}(i \omega_{m}) \hbar}{i (i \omega_{m})}$.  In the above equation, $\varepsilon_{n}=(2n+1)i\pi k_BT$ and $\omega_{m}=2mi\pi k_BT$ are the Matsubara energies, while $G_{\mathbf{k}}(i \varepsilon_{n})$ is the Matsubara Green function (in the $2\times 2$ matrix form). Taking into account the explicit form of $\hat{H}_{\mathbf{A}}^{\scriptstyle{E}}(i \omega_{m})$,
one can rewrite Eq.(\ref{2})  in the form
\begin{eqnarray}
\label{4}
j_{x}^{s_{z}} (i \omega_{m})= - \frac{1}{\beta}\frac{e E_{y}(i\omega_m) \hbar}{i (i \omega_{m})} \hspace{1.5cm} \nonumber \\
\times \sum_{\mathbf{k}, n} \mathrm{Tr}\left\{\hat{j}_{x}^{s_{z}} G_{\mathbf{k}}(i \varepsilon_{n} + i \omega_{m}) \hat{v}_{y} G_{\mathbf{k}}(i \varepsilon_{n}) \right\}.
\end{eqnarray}

The sum over Matsubara energies can be calculated by the method of contour integration,\cite{mahan}
\begin{eqnarray}
\label{5}
\frac{1}{\beta} \sum_{n} \hat{j}_{i}^{s_{\alpha}} G_{\mathbf{k}}(i \varepsilon_{n} + i \omega_{m}) \hat{v}_{j} G_{\mathbf{k}}(i \varepsilon_{n}) \hspace{0.8cm} \nonumber\\
= - \int_{\mathcal{C}}  \frac{dz}{2 \pi i} f(z) \hat{j}_{i}^{s_{\alpha}} G_{\mathbf{k}}(z + i \omega_{m}) \hat{v}_{j} G_{\mathbf{k}}(z) ,
\end{eqnarray}
where $\mathcal{C}$ is the appropriate contour of integration (for details see Ref.[\onlinecite{mahan}]).
As a result, one can write the frequency-dependent spin Hall conductivity in the form
\begin{eqnarray}
\label{6}
\sigma_{xy}^{s_{z}}(\omega) = \nonumber \\
- \frac{e \hbar}{\omega} {\mathrm{Tr}} \sum_{\mathbf{k}} \left[ \int \frac{d \varepsilon}{2 \pi} f(\varepsilon) \hat{j}_{x}^{s_{z}} G_{\mathbf{k}}^{R}(\varepsilon + \omega) \hat{v}_{y} [G_{\mathbf{k}}^{R}(\varepsilon) - G_{\mathbf{k}}^{A}(\varepsilon)]\right. \nonumber\\
+ \left. \int \frac{d \varepsilon}{2 \pi} f(\varepsilon + \omega)\hat{j}_{x}^{s_{z}} [G_{\mathbf{k}}^{R}(\varepsilon + \omega) - G_{\mathbf{k}}^{A}(\varepsilon + \omega)] \hat{v}_{y} G_{\mathbf{k}}^{A}(\varepsilon)\right],\nonumber\\
\end{eqnarray}
where $G_{\mathbf{k}}^{R}(\varepsilon)$ and $G_{\mathbf{k}}^{A}(\varepsilon)$ are the impurity-averaged retarded and advanced Green functions corresponding to the Hamiltonian (1), respectively,
while $f(\varepsilon )$ is the Fermi-Dirac distribution function.

It is convenient to write the retarded Green's function in the form
\begin{equation}
\label{9}
G_{\mathbf{k}}^{R}(\varepsilon) = G_{0}^{R}(\varepsilon)\, \sigma_{0} + G_{x}^{R}(\varepsilon)\, \sigma_{x} + G_{y}^{R}(\varepsilon)\, \sigma_{y},
\end{equation}
where
\begin{subequations}
\begin{align}
\label{9a}
G_{0}^{R}(\varepsilon) = \frac{1}{2} (G_{+}^{R}(\varepsilon) + G_{-}^{R}(\varepsilon)),\\
\label{9b}
G_{x}^{R}(\varepsilon) = \frac{1}{2} \sin(\phi) (G_{+}^{R}(\varepsilon) - G_{-}^{R}(\varepsilon)),\\
\label{9c}
G_{y}^{R}(\varepsilon) = - \frac{1}{2} \cos{\phi} (G_{+}^{R}(\varepsilon) - G_{-}^{R}(\varepsilon)),
\end{align}
\end{subequations}
and
\begin{eqnarray}
\label{10}
G_{\pm}^{R}(\varepsilon) = \frac{1}{\varepsilon + \mu - E_{\pm} + i \Gamma}.
\end{eqnarray}
Here, $\Gamma$ is the imaginary part of the self energy, related to the corresponding relaxation time $\tau$, $\Gamma =\hbar/2\tau$.
Note that $G_{0,x,y}^{R}(\varepsilon)$ as well as $G_{\pm}^{R}(\varepsilon)$ depend on $k$, which is not indicated explicitly. Similar form also holds for the advanced Green function. Equation (9) together with the formulas (10)-(12) will be used in section III to calculate the spin Hall conductivity first in the {\it bare bubble approximation} and then with the vertex correction included.

\begin{figure}
  \centering
  \includegraphics[width=0.6\columnwidth]{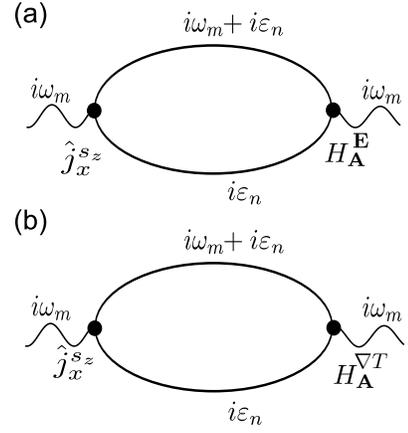}\\
  \caption{Loop diagram for (a) spin Hall and (b) spin Nernst conductivity in terms of the Matsubara Green functions.}
\end{figure}

\subsection{Spin Nernst effect}

Spin Nernst effect  differs from the spin Hall one by a driving force, which now is a temperature gradient instead of electric field (gradient of electrostatic potential). Therefore, SNE  is also frequently referred to as the thermally-induced spin Hall effect (or thermo-spin Hall effect). The formalism used in the case of electric field as a driving force can be also employed in the case of spin current induced by a temperature gradient. Based on the concept of {\it gravitational potential},~\cite{Luttinger,Strinati} one can define an auxiliary  time-dependent vector field of frequency $\omega/\hbar$, $\mathbf{A}(t)=\mathbf{A}(\omega)\exp (-i\omega t/\hbar) $  which is associated with
the heat current density operator $\hat{{\bf j}}^{h}$,
\begin{equation}
\hat{{\bf j}}^{h} = \frac{1}{2} \left[\hat{H}-\mu, \hat{\bf v} \right]_{+},
\end{equation}
so the perturbation term has the form similar to that given by Eq.~(2), i.e. $\hat{H}_{\mathbf{A}}^{\scriptstyle{\nabla T}}(t) = - \hat{{\bf j}}^{h} \cdot {\mathbf A}(t)$.\cite{Dyrdal2013,Tatara2015_1,Tatara2015_2} Here,  $\mu $ denotes the chemical potential.
This vector field is related to the temperature gradient {\it via}  the formula ${\mathbf A}(\omega)= \frac{\hbar}{i \omega} \left(- \frac{{\mathbf \nabla} T (\omega)}{T}\right)$.

When the temperature gradient is along the axis $y$, then the perturbation $\hat{H}_{\mathbf{A}}^{\scriptstyle{\nabla T}}(t)$ can be written in the form
\begin{equation}
\label{12}
\hat{H}_{\mathbf{A}}^{\scriptstyle{\nabla T}}(t) = - \hat{j}^{h}_{y} A_{y}(t).
\end{equation}
The spin current created by the temperature gradient can be then calculated by the Matsubara formalism from the formula similar to Eq.(\ref{2}),
\begin{eqnarray}
\label{11}
j_{x}^{s_{z}} (i \omega_{m})= \frac{1}{\beta}   \nonumber \\
\times \sum_{\mathbf{k}, n} \mathrm{Tr}\left\{\hat{j}_{x}^{s_{z}} G_{\mathbf{k}}(i \varepsilon_{n} + i \omega_{m}) \hat{H}_{\mathbf{A}}^{\scriptstyle{\nabla T}} (i \omega_{m})G_{\mathbf{k}}(i \varepsilon_{n}) \right\} ,
\end{eqnarray}
where now $\hat{H}_{\mathbf{A}}^{\scriptstyle{\nabla T}}(i \omega_{m})=-\hat{j}^{h}_{y} A_{y}(i\omega_m)$ and $A_{y}(i\omega_m)= \frac{\hbar}{i (i\omega_m)} \left(- \frac{\nabla_{y} T(i\omega_m)}{T}\right)$.
After summation over the Matsubara energies and defining the spin Nernst conductivity  by the relation  $\alpha_{xy}^{s_{z}} (\omega) = j_{x}^{s_{z}}(\omega)/(- \nabla_{y} T(\omega))$, one finds the following expression for $\alpha_{xy}^{s_{z}}(\omega)$:
\begin{eqnarray}
\label{13}
\alpha_{xy}^{s_{z}} (\omega)= \nonumber \\
- \frac{\hbar}{\omega} \frac{1}{T} {\mathrm{Tr}} \sum_{\mathbf{k}} \left[ \int \frac{d \varepsilon}{2 \pi} f(\varepsilon) \hat{j}_{x}^{s_{z}} G_{\mathbf{k}}^{R}(\varepsilon + \omega) \hat{j}_{y}^{h} [G_{\mathbf{k}}^{R}(\varepsilon) - G_{\mathbf{k}}^{A}(\varepsilon)] \right. \nonumber\\
+ \left.  \int \frac{d \varepsilon}{2 \pi} f(\varepsilon + \omega) \hat{j}_{x}^{s_{z}} [G_{\mathbf{k}}^{R}(\varepsilon + \omega) - G_{\mathbf{k}}^{A}(\varepsilon + \omega)] \hat{j}_{y}^{h}G_{\mathbf{k}}^{A}(\varepsilon) \right].\nonumber\\
\end{eqnarray}

It is convenient to write the operator of  heat current density in the above equation, $\hat{j}_{y}^{h}$, in the form
\begin{eqnarray}
\label{14}
\hat{j}_{y}^{h} = j_{y, 0}^{h} \sigma_{0} + j_{y, x}^{h} \sigma_{x} + j_{y, y}^{h} \sigma_{y},
\end{eqnarray}
where
\begin{subequations}
\begin{align}
\label{15a}
j_{y, 0}^{h} = \frac{\hbar}{m} k (\varepsilon_{k} - \mu) \sin(\phi) + \frac{\alpha^{2}}{\hbar} k \sin(\phi),\\
\label{15b}
j_{y, x}^{h} = \frac{\alpha}{\hbar} (\varepsilon_{k} - \mu) + \frac{\alpha \hbar}{m} k^{2} \sin^{2}(\phi),\\
\label{15c}
j_{y, y}^{h} = -  \frac{\alpha \hbar}{m} k^{2} \sin(\phi) \cos(\phi).
\end{align}
\end{subequations}
Like in the case of SHE, Eq.(16) together with the formulas (17) and (18) will be used in Section IV to calculate the spin Nernst conductivity.

%----------------------------------
\section{Spin Hall conductivity}
%----------------------------------

In this section we use the general formulas derived above to calculate the spin Hall conductivity.  At the beginning we calculate this conductivity in the approximation corresponding to the \textit{bare bubble} diagram. In such a diagram the two Green functions are averaged over the impurity positions and electron scattering is taken into account  {\it via} the self-energy. Then we include vertex correction and show that the spin Hall conductivity vanishes at arbitrary temperature.

%======================================================================
\subsection{Bare bubble approximation}
%======================================================================

First, we rewrite Eq.~(9) for the spin Hall conductivity in the form
\begin{eqnarray}
\label{16}
\sigma_{xy}^{s_{z}} = - \frac{e \hbar}{\omega} \int \frac{dk k}{(2 \pi)^{2}} \left[ \int \frac{d \varepsilon}{2 \pi} f(\varepsilon) \mathcal{T}_{1}(\varepsilon + \omega, \varepsilon)\right. \nonumber\\ + \left.\int \frac{d \varepsilon}{2 \pi} f(\varepsilon)\mathcal{T}_{2}(\varepsilon , \varepsilon - \omega)\right],
\end{eqnarray}
where $\mathcal{T}_{1}(\varepsilon + \omega, \varepsilon)$ and $\mathcal{T}_{2}(\varepsilon , \varepsilon - \omega)$ are defined by  the following formulas:
\begin{eqnarray}
\label{17}
\mathcal{T}_{1}(\varepsilon + \omega, \varepsilon) \nonumber \\
=\int d\phi\, {\mathrm{Tr}}\left\{\hat{j}_{x}^{s_{z}} G_{\mathbf{k}}^{R}(\varepsilon + \omega) \hat{v}_{y} [G_{\mathbf{k}}^{R}(\varepsilon) - G_{\mathbf{k}}^{A}(\varepsilon)]\right\}, \\ \label{15}
\mathcal{T}_{2}(\varepsilon , \varepsilon - \omega) \nonumber \\
= \int d\phi\, {\mathrm{Tr}}\left\{ \hat{j}_{x}^{s_{z}} [G_{\mathbf{k}}^{R}(\varepsilon) - G_{\mathbf{k}}^{A}(\varepsilon)] \hat{v}_{y} G_{\mathbf{k}}^{A}(\varepsilon - \omega)\right\}.
\end{eqnarray}
Note, $\mathcal{T}_{1}(\varepsilon + \omega, \varepsilon)$ and $\mathcal{T}_{2}(\varepsilon , \varepsilon - \omega)$ depend on $k$, which for notation clarity  is not indicated explicitly.
Taking now into account explicit forms of the spin current density and electron velocity operators, as well as making use of Eqs. (10)-(12)  and  integrating over the angle $\phi$, one can write $\mathcal{T}_{1}$ and $\mathcal{T}_{2}$ as
\begin{eqnarray}
\label{18}
\mathcal{T}_{1}(\varepsilon + \omega, \varepsilon) \nonumber \\
= i \alpha \frac{\hbar k}{2 m} \pi \left[  G^{R}_{-}(\varepsilon + \omega) G^{A}_{+}(\varepsilon) -  G^{R}_{-}(\varepsilon + \omega) G^{R}_{+}(\varepsilon) \right.\nonumber\\
\left.  - G^{R}_{+}(\varepsilon + \omega) G^{A}_{-}(\varepsilon) + G^{R}_{+}(\varepsilon + \omega) G^{R}_{-}(\varepsilon) \right],
\end{eqnarray}
\begin{eqnarray}
\label{19}
\mathcal{T}_{2}(\varepsilon, \varepsilon - \omega) \nonumber \\
 = i \alpha \frac{\hbar k}{2 m} \pi \left[  G^{
A}_{-}(\varepsilon) G^{A}_{+}(\varepsilon - \omega)-  G^{A}_{+}(\varepsilon) G^{A}_{-}(\varepsilon - \omega) \right.\nonumber\\
\left.  - G^{R}_{-}(\varepsilon) G^{A}_{+}(\varepsilon - \omega) + G^{R}_{+}(\varepsilon) G^{A}_{-}(\varepsilon - \omega) \right].
\end{eqnarray}
Then, upon  integrating over $\varepsilon$ and taking the limit $\omega \rightarrow 0$ (see Appendix A for details) we find
\begin{eqnarray}
\label{20}
\sigma_{xy}^{s_z} = - \frac{e \hbar^{2}}{16 \pi m \alpha} \int dk [f(E_{+}) - f(E_{-})]\nonumber\\
+ \frac{e \hbar^{2}}{16 \pi m} \int dk k \frac{f'(E_{+}) + f'(E_{-})}{1 + (\alpha k /\Gamma)^{2}},
\end{eqnarray}
where $f^\prime (\varepsilon) \equiv \partial f(\varepsilon) /\partial \varepsilon $.
The above equation is our final result for the temperature dependent spin Hall conductivity in the {\it bare} bubble approximation, which will be used in numerical
calculations. Some analytical results can be obtained in the zero-temperature limit, as shown below.
To do this let us rewrite Eq.~(\ref{20}) in the form
\begin{equation}
\label{21}
\sigma_{xy}^{s_{z}} = -  \frac{e \hbar^{2}}{16 \pi m \alpha} \mathcal{I}_{1} + \frac{e \hbar^{2}}{16 \pi m} \mathcal{I}_{2},
\end{equation}
where
\begin{eqnarray}
\mathcal{I}_{1} = \int dk [f(E_{+}) - f(E_{-})], \\
\mathcal{I}_{2} = \int dk k \frac{f'(E_{+}) + f'(E_{-})}{1 + (\alpha k /\Gamma )^{2}}.
\end{eqnarray}

In the zero temperature limit both $\mathcal{I}_{1}$ and $\mathcal{I}_{2}$ can be evaluated analytically.
In the case of $\mu =\mu_0>0$ (here $\mu_0$ is the chemical potential at $T=0$, i.e. the Fermi energy) one finds
\begin{eqnarray}
\label{22}
\mathcal{I}_{1} =  - \frac{2 m \alpha}{\hbar^{2}},
\end{eqnarray}
and
\begin{eqnarray}
\label{23}
\mathcal{I}_{2} =  - \frac{m}{\sqrt{2 m \mu_0 \hbar^{2} + m^{2} \alpha^{2}}} \nonumber \\
\times \left[ \frac{k_{+}}{1 + (\alpha k_{+} /\Gamma)^{2}} + \frac{k_{-}}{1 + (\alpha k_{-} / \Gamma)^{2}}\right],
\end{eqnarray}
where $k_{\pm} = \mp \frac{m \alpha}{\hbar^{2}} + \frac{1}{\hbar^{2}} \sqrt{2 m \mu_0 \hbar^{2} + m^{2} \alpha^{2}}$ are the Fermi wavevectors corresponding to the two electronic subbands $E_\pm$.
Then, upon  substituting Eqs (\ref{22}) and (\ref{23}) into Eq.~(\ref{21}) we arrive at
\begin{eqnarray}
\label{24}
\sigma_{xy}^{s_{z}} = \frac{e}{8 \pi}- \frac{e \hbar^{2}}{16 \pi \sqrt{2 m \mu_0 \hbar^2 + m^{2} \alpha^{2}}}\hspace{2cm}\nonumber\\ \times \left[ \frac{k_{+}}{1 + (\alpha k_{+}/ \Gamma)^{2}} + \frac{k_{-}}{1 + (\alpha k_{-}/\Gamma)^{2}}\right].
\end{eqnarray}
This formula may be also rewritten in the equivalent, form which is more convenient for interpretation,
\begin{eqnarray}
\label{25}
\sigma_{xy}^{s_{z}} = \frac{e}{8 \pi}- \frac{e}{16 \pi} \left[ \frac{1}{1 + (\alpha k_{+} /\Gamma)^{2}} + \frac{1}{1 + (\alpha k_{-}/
\Gamma)^{2}}\right]\nonumber\\ + \frac{e}{16 \pi} \frac{n^{\ast}}{n}\left[ \frac{1}{1 + (\alpha k_+/\Gamma)^{2}} - \frac{1}{1 + (\alpha k_{-}
/\Gamma)^{2}}\right],
\end{eqnarray}
where $n=\frac{m \alpha}{\pi \hbar^{4}} \sqrt{m^{2} \alpha^{2} + 2 m \mu_0 \hbar^{2}}$ and $n^{\ast} = \frac{m^{2} \alpha^{2}}{\pi \hbar^{4}}$ are the electron concentrations corresponding to the Fermi energy $\mu_0$ and to $\mu_0=0$, respectively.
In the limit of small concentration of impurities (infinitely long relaxation time, or equivalently $\Gamma\to 0$), the above equation gives the well known universal intrinsic value of the spin Hall conductivity, i.e. $\sigma_{xy}^{s_{z}} = \sigma_{xy}^{s_{z},0} = \frac{e}{8 \pi}$. On the other side, when the relaxation time is finite (i.e. we are beyond the ballistic limit), and if we assume rather weak spin-orbit interaction, then for $\frac{\alpha^{2} m}{2 \hbar^{2}} \ll \mu_0$ one may assume $k_{+} \approx k_{-} \approx k_{0} = \sqrt{2m\mu_0 }/\hbar$ to recover the result derived by Dimitrova,\cite{Dimitrova2005}
\begin{eqnarray}
\label{26}
\sigma_{xy}^{s_{z}} = \frac{e}{8 \pi} \left( 1 - \frac{1}{1 + (\alpha k_{0} /\Gamma)^{2}}\right).
\end{eqnarray}

Let us consider validity of the above results.
Detailed calculations of the zero-temperature self energy (see Appendix B) show that
$\Gamma$ is finite and constant for $\mu_0 >0$ (see also Ref.~\onlinecite{Borosco}),
\begin{equation}
\Gamma = \pi n_{i} V^{2} \frac{m}{2 \pi \hbar^{2}}\equiv \Gamma_0,
\end{equation}
where $n_i$ is the impurity concentration and $V$ is the strength of the point-like scattering potential of impurities.
For $\mu_0 <0$, in turn, $\Gamma$ depends on energy and takes the form
\begin{equation}
\Gamma = \pi n_{i} V^{2} \frac{m}{2 \pi \hbar^{2}} \frac{n^{\ast}}{n} = \Gamma_0 \frac{n^{\ast}}{n}.
\end{equation}
Thus, $\Gamma \to \infty$ when $\mu_0$ approaches the bottom of the lower band ($n\to 0$). Accordingly, the role of disorder effectively increases when
the limit $\mu_0 = \mu_{\mathrm{min}} = - \frac{m\alpha^{2}}{2 \hbar^{2}}$ is approached, and the localization Ioffe-Regel criterion is obeyed for a certain $\mu_0$ above $\mu_{\mathrm{min}}$. Thus we need to find the limit for chemical potential, $\mu_{\mathrm{loc}}$, below which the states become localized and the conductivity is suppressed to zero.

The Ioffe-Regel localization criterion can be written as
\begin{equation}
\mu_{\mathrm{loc}} - \mu_{\mathrm{min}} \simeq \Gamma(\mu_{\mathrm{loc}}).
\end{equation}
When $n_{i}V^{2}/\alpha^{2} <1$, then $\mu_{\mathrm{loc}}<0$ and
\begin{equation}
\mu_{\mathrm{loc}} \simeq  - \frac{m \alpha^{2}}{2 \hbar^{2}} \left(1 - \frac{n_{i}V^{2}}{\alpha^{2}}\right).
\end{equation}
For this particular value of the chemical potential one finds
$n^{\ast}/n(\mu_{\mathrm{loc}}) = \alpha /\sqrt{n_{i} V^{2}}$, so
\begin{equation}
\Gamma =\Gamma_0 \frac{\alpha}{\sqrt{n_{i} V^{2}}}.
\end{equation}
In turn, when $n_{i}V^{2}/\alpha^{2} \ge 1$, then
$\mu_{\mathrm{loc}} \ge 0$, and
\begin{equation}
\mu_{\mathrm{loc}} \simeq   \Gamma_0+\mu_{\mathrm{min}},
\end{equation}
while $\Gamma =\Gamma_0$. Since the Rashba parameter $\alpha$ is usually small, then $\mu_{\mathrm{loc}} > 0$. Therefore, further considerations are limited to  $\mu_0 \ge 0$, which is a physically relevant regime.

\begin{figure}
  \centering
 \includegraphics[width=0.97\columnwidth]{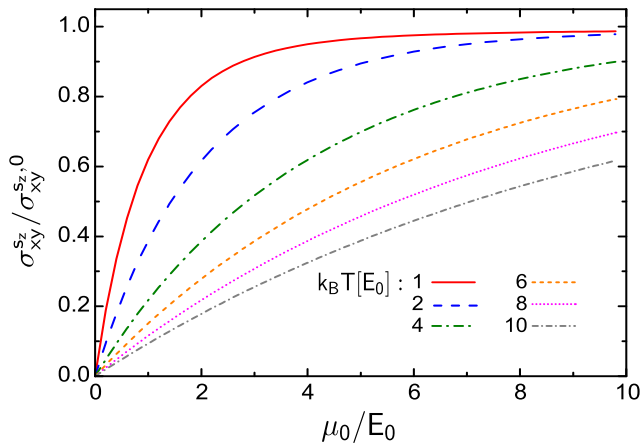}
  \caption{(Color online) Spin Hall conductivity in the {\it bare bubble} approximation as a function of $\mu_0/E_0$, where $\mu_0$ is the  Fermi energy at $T=0$, while
$E_0= \frac{m \alpha^{2}}{\hbar^{2}}$. Different curves correspond to indicated values of temperature. For simplicity we put
 $m = 1$, $\hbar = 1$, $k_{B} = 1$, and assumed $\alpha =1$ and $\Gamma = 0.5 E_{0}$, so the localization threshold is $\mu_{\rm loc}=0$. }\label{SHE_Fig1_2}
\end{figure}
Consider now some numerical results for the spin Hall conductivity in the \textit{bare bubble} approximation.
In Fig.2 we show the spin Hall conductivity as a function of $\mu_0/E_0$ for
different temperatures, where
$E_0= m \alpha^{2}/\hbar^{2}$. We consider the situation when the particle number is constant, so the chemical potential at a nonzero  temperature is adapted following the formula
$ \mu = k_{B} T \ln[\exp[\mu_0/(k_{B} T)] - 1]$.\cite{Vignale2008}
For simplicity we used the units with  $m = 1$, $\hbar = 1$, and $k_{B} = 1$. Apart from this we assumed  $\alpha =1$ and $\Gamma = 0.5 E_{0}$. This corresponds to the localization threshold $\mu_{\rm loc}=0$. As follows from this figure,  the spin Hall conductivity increases with increasing $\mu_0/E_0$,
and for large values of $\mu_0/E_0$ it tends to a constant and universal value $e/8\pi$. The rate of this increase is larger for smaller $T$ and smaller for higher temperatures.

%==========================================================
\subsection{Vertex correction}
%==========================================================

Now we consider the role of impurities taking into account correction to the spin Hall  conductivity
due to renormalization of the spin current vertex.
It is well known that vertex corrections can have a significant influence on the spin current induced {\it via} the spin Hall effect. In the case of two-dimensional electron gas with Rashba spin-orbit interaction such a vertex correction can totally suppress the spin Hall conductivity at $T=0$.\cite{Dimitrova2005,Inoue2004,ChalaevLoss2004,Mishchenko2004} It has been predicted that this correction can suppress the conductivity also at arbitrary $T$.\cite{Dimitrova2005,Erlingsson2005,Schliemann2006} Here we consider this correction at finite temperatures within the Matsubara formalism.
\begin{figure}[t]
  \centering
  \includegraphics[width=0.9\columnwidth]{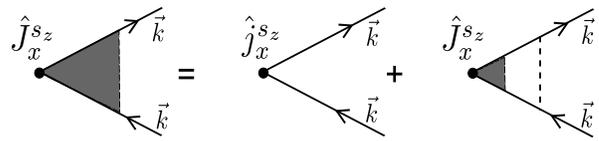}
  \caption{Graphical equation for the renormalized spin current vertex. }
\end{figure}

The equation describing the renormalized vertex corresponding to the spin current takes the form
\begin{equation}
\label{vertexEq}
\hat{J}_{x}^{s_{z}} = \hat{j}_{x}^{s_{z}} + \int \frac{d^{2}\mathbf{k}}{(2\pi)^{2}} V^{2} G_{\mathbf{k}}^{A}(\varepsilon) \hat{J}_{x}^{s_{z}}G_{\mathbf{k}}^{R}(\varepsilon + \omega),
\end{equation}
where the spin current operator $\hat{j}_{x}^{s_{z}}$ is given by Eq.~(5), and
the vertex correction (second term on the right-hand side) has to be taken at the Fermi level, $\varepsilon = 0$, and in the limit $\omega \rightarrow 0$.
The above equation is presented graphically in Fig.3. Note, the vertex correction term in Eq.~(\ref{vertexEq}) with two retarded or two advanced Green's functions is equal to zero in the limit of $\omega \rightarrow 0$.

We look for a solution of Eq.(\ref{vertexEq}) in the following form:
\begin{equation}
\hat{J}_{x}^{s_{z}} = a \sigma_{0} + b \sigma_{x} + c \sigma_{y} + d k_{x} \sigma_{z}
\end{equation}
where $a$, $b$, $c$ and $d$ are coefficients that need to be determined.
Detail calculations (see Appendix C) show that the renormalized spin current vertex can be written in the form\cite{Tse}
\begin{equation}
\hat{J}_{x}^{s_{z}} = \hat{j}_{x}^{s_{z}} + \hat{\delta j}_{x}^{s_{z}},
\end{equation}
where the correction $\hat{\delta j}_{x}^{s_{z}}$ is
\begin{equation}
\hat{\delta j}_{x}^{s_{z}} = b\, \sigma_{x},
\end{equation}
with the
coefficient $b$ given by
\begin{equation}
\label{b1}
b = \frac{\hbar^{2}}{2 m} \frac{\Gamma}{\alpha}.
\end{equation}
We recall, that $\Gamma$  is constant for $\mu_0 >0$,
see Eq.~(33).

The total spin Hall conductivity $\sigma_{xy}^{s_{z}\, {\mathrm{tot}}}$ is a sum of the contribution $\sigma_{xy}^{s_{z}}$ from the {\it bare bubble} diagram and the contribution $\delta \sigma_{xy}^{s_{z}}$ due to vertex correction,
\begin{equation}
\sigma_{xy}^{s_{z}\, {\mathrm{tot}}} = \sigma_{xy}^{s_{z}} + \delta \sigma_{xy}^{s_{z}}.
\end{equation}
The {\it bare bubble} contribution, $\sigma_{xy}^{s_{z}}$, was calculated and analyzed in Sec.IIIA, see Eq.~(24).
In turn,
the contribution $\delta \sigma_{xy}^{s_{z}}$ to the spin Hall conductivity due to the vertex correction
is determined by the formula
\begin{eqnarray}
\delta \sigma_{xy}^{s_{z}} =  \frac{e \hbar}{\omega} {\mathrm{Tr}} \int \frac{d^{2} \mathbf{k}}{(2 \pi)^{2}} \left[ \int \frac{d \varepsilon}{2 \pi} f(\varepsilon) \hat{\delta j}_{x}^{s_{z}} G_{\mathbf{k}}^{R}(\varepsilon + \omega) \hat{v}_{y} G_{\mathbf{k}}^{A}(\varepsilon)\right.\nonumber\\
- \left. \int \frac{d \varepsilon}{2 \pi} f(\varepsilon) \hat{\delta j}_{x}^{s_{z}} G_{\mathbf{k}}^{R}(\varepsilon) \hat{v}_{y} G_{\mathbf{k}}^{A}(\varepsilon - \omega)\right].\hspace{1cm}
\end{eqnarray}

Upon integrating over $\varepsilon$, (for details see Appendix C) and taking into account the explicit form of the parameter $b$ [see Eq.(43)], this formula leads to the following expression for the vertex correction to the spin Hall conductivity:
\begin{eqnarray}
\delta \sigma_{xy}^{s_{z}} = - \frac{e \hbar^{2}}{4 m} \int \frac{dk k}{4 \pi} [f'(E_{+}) + f'(E_{-})]\nonumber\\
- \frac{e \hbar^{2}}{4 m} \int \frac{dk k}{4 \pi} \frac{f'(E_{+}) + f'(E_{-})}{1 + \left(\frac{\alpha}{\Gamma} k\right)^{2}}\nonumber\\
- \frac{e \hbar^{4}}{4 m^{2} \alpha} \int \frac{dk k}{4 \pi} k [f'(E_{+}) - f'(E_{-})].
\end{eqnarray}

In the zero-temperature limit and for $\mu_0>0$,  Eq.(46) gives
\begin{eqnarray}
\delta \sigma_{xy}^{s_{z}} = - \frac{e}{8 \pi} + \frac{e \hbar^{2}}{16 \pi \sqrt{m^{2} \alpha^{2} + 2 m \mu_0 \hbar^{2}}}\hspace{1cm}\nonumber\\
\times \left[ \frac{k_{+}}{1 + \left( \frac{\alpha}{\Gamma} k_{+}\right)^{2}} + \frac{k_{-}}{1 + \left( \frac{\alpha}{\Gamma} k_{-}\right)^{2}}\right].
\end{eqnarray}
Taking into account Eq.(30), one immediately concludes that
$\sigma_{xy}^{s_{z}\, {\mathrm{tot.}}} = 0$ at $T=0$.\cite{Dimitrova2005,Inoue2004,ChalaevLoss2004,Mishchenko2004}

Consider now the case of an arbitrary temperature. Taking into account Eq.(24) and Eq.(46), one finds the total spin Hall conductivity in the form
\begin{eqnarray}
\sigma_{xy}^{s_{z}\, {\mathrm{tot.}}} = - \frac{e \hbar^{2}}{16 \pi m \alpha} \int dk [f(E_{+}) - f(E_{-})]\nonumber\\
- \frac{e \hbar^{2}}{16 \pi m} \int dk k [f'(E_{+}) + f'(E_{-})]\nonumber\\
- \frac{e \hbar^{4}}{16 \pi m^{2} \alpha} \int dk k^{2} [f'(E_{+}) - f'(E_{-})].
\end{eqnarray}
This rather long formula can be presented as an integral over $k$ of a full derivative,
\begin{eqnarray}
\label{66a}
\sigma_{xy}^{s_{z}\, {\mathrm{tot.}}} = - \frac{e \hbar^{2}}{16 \pi m \alpha} \int dk \,
\frac{d }{d k}\Big\{ k\, [f(E_{+}) - f(E_{-})]\Big\} ,\hskip0.5cm
\end{eqnarray}
which gives zero after integration over $k$ from $0$ to $\infty$,
\begin{equation}
\sigma_{xy}^{s_{z}\, {\mathrm{tot.}}} = 0.
\end{equation}
Thus, the total  spin Hall conductivity (including the vertex correction) vanishes at any temperature, in agreement with
earlier predictions.\cite{Dimitrova2005,Erlingsson2005,Schliemann2006}

%----------------------------------
\section{Spin Nernst conductivity}
%----------------------------------

In this section we consider the spin Nernst conductivity using the general formulas derived in section 2B. Accordingly, we calculate this conductivity first in the {\it bare bubble} approximation, and then derive the vertex correction. To get physically correct results, we also calculate the correction due to orbital spin-resolved magnetization. The total spin Nernst conductivity vanishes then at $T=0$, but is finite at nonzero $T$.

%======================================================================
\subsection{Bare bubble approximation}
%======================================================================

Equation (16) for the spin Nernst conductivity can be written in the form
\begin{eqnarray}
\label{32}
\alpha_{xy}^{s_{z}} = - \frac{e \hbar}{\omega} \int \frac{dk k}{(2 \pi)^{2}} (\varepsilon_{k} -\mu)\left[ \int \frac{d \varepsilon}{2 \pi} f(\varepsilon) \mathcal{T}_{1}(\varepsilon + \omega, \varepsilon)\right. \nonumber\\ + \left.\int \frac{d \varepsilon}{2 \pi} f(\varepsilon)\mathcal{T}_{2}(\varepsilon , \varepsilon - \omega)\right],\hspace{0.8cm}
\end{eqnarray}
where $\mathcal{T}_{1}$ and $\mathcal{T}_{2}$ are given by Eqs (22) and (23). Following the procedure used in the case of spin Hall conductivity (see also Appendix A), we find
\begin{eqnarray}
\label{33}
\alpha_{xy}^{s_{z}} = - \frac{\hbar^{2}}{16 \pi m \alpha} \frac{1}{T} \int dk  (\varepsilon_{k} - \mu) [f(E_{+}) - f(E_{-})]\nonumber\\  + \frac{\hbar^{2}}{16 \pi m} \frac{1}{T} \int dk k (\varepsilon_{k} - \mu) \frac{f'(E_{+}) + f'(E_{-})}{1 + (\alpha k /\Gamma )^{2}}.\;\;
\end{eqnarray}
The above expression is a general formula for the spin Nernst conductivity in the {\it bare bubble} approximation, which is valid for arbitrary temperature.
Now, we consider in more details the low-temperature regime, where some analytical results can be obtained.
To do this, let us rewrite Eq.(\ref{33}) in the following form:
\begin{equation}
\alpha_{xy}^{s_{z}} = - \frac{\hbar^{2}}{16 \pi m \alpha} \frac{1}{T} \mathcal{I}_{3}  + \frac{\hbar^{2}}{16 \pi m} \frac{1}{T} \mathcal{I}_{4},
\end{equation}
where
\begin{eqnarray}
\mathcal{I}_{3} = \int dk  (\varepsilon_{k} - \mu) [f(E_{+}) - f(E_{-})], \\
\mathcal{I}_{4} = \int dk k (\varepsilon_{k} - \mu) \frac{f'(E_{+}) + f'(E_{-})}{1 + (\alpha k /\Gamma )^{2}}.
\end{eqnarray}

In the low-temperature regime one can replace $\mathcal{I}_{3}$ and $\mathcal{I}_{4}$ by the corresponding zero-temperature values, which can be found analytically. From Eqs.(54) and (55) and for $\mu_0 >0$ one finds
\begin{eqnarray}
\mathcal{I}_{3} =
- \frac{4}{3} \frac{m^{2} \alpha^{3}}{\hbar^{4}},
\end{eqnarray}
and
\begin{eqnarray}
\mathcal{I}_{4} =  -\frac{m}{\sqrt{m^{2} \alpha^{2} + 2 m \mu_0 \hbar^{2}}} \nonumber \\
\times \left[ \frac{k_{+} (\frac{\hbar^{2}k_{+}^{2}}{2 m} - \mu_0)}{1 + \left( \alpha k_{+}/\Gamma\right)^{2}} + \frac{k_{-} (\frac{\hbar^{2}k_{-}^{2}}{2 m} - \mu_0)}{1 + \left( \alpha k_{-}/\Gamma \right)^{2}}\right].
\end{eqnarray}
Thus, the spin Nernst conductivity takes then the form
\begin{eqnarray}
\alpha_{xy}^{s_{z}} = \frac{1}{T} \frac{m \alpha^{2}}{12 \pi \hbar^{2}} - \frac{1}{T} \frac{\hbar^{2}}{16 \pi \sqrt{m^{2} \alpha^{2} + 2 m \mu_0 \hbar^{2}}} \hspace{0.7cm}\nonumber\\
\times \left[ \frac{k_{+} (\frac{\hbar^{2}k_{+}^{2}}{2 m} - \mu_0)}{1 + \left( \alpha k_{+}/\Gamma\right)^{2}} + \frac{k_{-} (\frac{\hbar^{2}k_{-}^{2}}{2 m} - \mu_0)}{1 + \left( \alpha k_{-}/\Gamma \right)^{2}}\right],
\end{eqnarray}
or equivalently
\begin{eqnarray}
\alpha_{xy}^{s_{z}} = \frac{1}{T}\frac{m \alpha^{2}}{12 \pi \hbar^{2}} \hspace{4.5cm}\nonumber\\
- \frac{m \alpha^{2}}{4 \pi \hbar^{2}} \frac{1}{T} \frac{1}{[1 + (\alpha k_{+} /\Gamma )^{2}][1 + (\alpha k_{-} /\Gamma )^{2}]}.
\end{eqnarray}
In the limit of $\Gamma \rightarrow 0$, the spin Nernst conductivity
is given only by the first term in the above equations. This result is consistent with that obtained by Ma.~\cite{Ma}
For systems with a small $\alpha$ and for large values of the chemical potential we get
\begin{equation}
\alpha_{xy}^{s_{z}} = \frac{1}{T} \frac{m \alpha^{2}}{12 \pi \hbar^{2}} \left[ 1 - \frac{3}{1 + 2 k_{0}^{2} \alpha^{2}/\Gamma^2 }\right].
\end{equation}

%==========================================================
\subsection{Vertex correction}
%==========================================================

Similarly to the spin Hall conductivity, the renormalized spin Nernst conductivity $\alpha_{xy}^{s_{z}\, {\mathrm{ren}}}$ is a sum of the contribution $\alpha_{xy}^{s_{z}}$ from the {\it bare bubble} diagram and the contribution $\delta \alpha_{xy}^{s_{z}}$ due to the vertex correction,
\begin{equation}
\alpha_{xy}^{s_{z}\, {\mathrm{ren}}} = \alpha_{xy}^{s_{z}} + \delta \alpha_{xy}^{s_{z}}.
\end{equation}
The bare bubble contribution, $\alpha_{xy}^{s_{z}}$, was calculated above, see Eq.~(52).
In turn, the vertex correction $\delta \alpha_{xy}^{s_{z}}$ can be calculated from the formula
\begin{eqnarray}
\delta \alpha_{xy}^{s_{z}} =  \frac{\hbar}{\omega} \frac{1}{T} {\mathrm{Tr}} \int \frac{d^{2} \mathbf{k}}{(2 \pi)^{2}} \left[ \int \frac{d \varepsilon}{2 \pi} f(\varepsilon) \hat{\delta j}_{x}^{s_{z}} G_{\mathbf{k}}^{R}(\varepsilon + \omega) \hat{j}_{y}^{h} G_{\mathbf{k}}^{A}(\varepsilon)\right.\nonumber\\
- \left. \int \frac{d \varepsilon}{2 \pi} f(\varepsilon) \hat{\delta j}_{x}^{s_{z}} G_{\mathbf{k}}^{R}(\varepsilon) \hat{j}_{y}^{h} G_{\mathbf{k}}^{A}(\varepsilon - \omega)\right].\hskip0.5cm
\end{eqnarray}

Following the procedure used in the case of spin Hall conductivity (see also Appendix C), and taking into account the explicit form of the
renormalization parameters, one finds
\begin{eqnarray}
\delta \alpha_{xy}^{s_{z}} = - \frac{\hbar^{2}}{4 m} \frac{1}{T} \int \frac{dk}{4 \pi} k (\varepsilon_{k} - \mu) [f'(E_{+}) + f'(E_{-})]\nonumber\\
- \frac{\hbar^{2}}{4 m} \frac{1}{T} \int \frac{dk}{4\pi} k (\varepsilon_{k} - \mu) \frac{f'(E_{+}) + f'(E_{-})}{1 + \left( \frac{\alpha}{\Gamma} k\right)^{2}}\nonumber\\
- \frac{\hbar^{4}}{4 m^{2} \alpha} \frac{1}{T} \int \frac{dk}{4 \pi} k^{2}(\varepsilon_{k} - \mu) [f'(E_{+}) - f'(E_{-})]\nonumber\\
- \frac{\hbar^{4}}{4 m^{2}} \frac{1}{T} \int \frac{dk}{4\pi} k^{3} [f'(E_{+}) + f'(E_{-})]\nonumber\\
- \frac{\hbar^{2}}{4 m} \frac{\alpha}{T} \int \frac{dk}{4\pi} k^{2} [f'(E_{+}) - f'(E_{-})].\hskip0.4cm
\end{eqnarray}

In the low-temperature regime, one can use the zero-temperature limit for the distribution function in the above equation, and for $\mu_0>0$ one finds
\begin{eqnarray}
\delta \alpha_{xy}^{s_{z}} = \frac{1}{T} \frac{\hbar^{2}}{16 \pi \sqrt{m^{2} \alpha^{2} + 2 m \mu_0 \hbar^{2}}}\hspace{1.7cm}\nonumber\\
\times \left[ \frac{k_{+} (\frac{\hbar^{2} k_{+}^{2}}{2 m} - \mu_0)}{1 + \left( \frac{\alpha}{\Gamma} k_{+}\right)^{2}} + \frac{k_{-} (\frac{\hbar^{2} k_{-}^{2}}{2 m} - \mu_0)}{1 + \left( \frac{\alpha}{\Gamma} k_{-}\right)^{2}}\right].
\end{eqnarray}
Taking into account Eq.(58), the renormalized spin Nernst conductivity can be written in the form
\begin{equation}
\alpha_{xy}^{s_{z}\,{\mathrm{ren}}} = \frac{1}{T} \frac{m \alpha^{2}}{12 \pi \hbar^{2}}\equiv \alpha_{xy}^{s_{z}\,{0}}.
\end{equation}

 For higher temperatures, the renormalized  spin Nernst conductivity, including the vertex correction, is given by the following expression:
\begin{eqnarray}
\alpha_{xy}^{s_{z}\,{\mathrm{ren}}} = - \frac{1}{T} \frac{\hbar^{2}}{16 \pi m \alpha} \int dk (\varepsilon_{k} - \mu) [f(E_{+}) - f(E_{-})]\nonumber\\
 - \frac{\hbar^{2}}{16 \pi m} \frac{1}{T} \int dk k (\varepsilon_{k} - \mu) [f'(E_{+}) + f'(E_{-})]\nonumber\\
- \frac{\hbar^{4}}{16 \pi m^{2} \alpha} \frac{1}{T} \int dk k^{2}(\varepsilon_{k} - \mu) [f'(E_{+}) - f'(E_{-})]\nonumber\\
- \frac{\hbar^{4}}{16 \pi m^{2}} \frac{1}{T} \int dk k^{3} [f'(E_{+}) + f'(E_{-})]\nonumber\\
- \frac{\hbar^{2}}{16 \pi m} \frac{\alpha}{T} \int dk k^{2} [f'(E_{+}) - f'(E_{-})].\hspace{1.0cm}
\end{eqnarray}
This formula can be rewritten in the following form:
\begin{eqnarray}
\alpha_{xy}^{s_{z}\,{\mathrm{ren}}} = - \frac{1}{T} \frac{\hbar^{2}}{16 \pi m} \left[
\int dk \frac{d}{d k} \Big\{k^{2} [f(E_{+}) + f(E_{-})] \Big\}
\right.\nonumber\\
 +\left. \frac{1}{\alpha} \int dk \frac{d}{d k} \Big\{ k (\varepsilon_{k} - \mu) [f(E_{+}) - f(E_{-})]\Big\}\right]\hspace{1.1cm}\nonumber\\
 + \frac{1}{T} \frac{\hbar^{2}}{8 \pi m \alpha} \int dk [E_{+} f(E_{+}) - E_{-} f(E_{-})]. \hspace{0.4cm}
\end{eqnarray}
The first two terms in this expression  give zero after integration over $k$ from $0$ to $\infty$. The final formula is then given only by the last term in Eq.(67):
\begin{equation}
\alpha_{xy}^{s_{z}\,{\mathrm{ren}}} =  \frac{1}{T} \frac{\hbar^{2}}{8 \pi m \alpha} \int dk [E_{+} f(E_{+}) - E_{-} f(E_{-})].
\end{equation}

From the above derived formulas follows that the renormalized spin Nernst conductivity (as well as that in the {\it bare bubble} approximation) is divergent in the limit of $T\to 0$, which is nonphysical.
Similar problem was also encountered in the case of Nernst effect,\cite{obraztsov,gusynin}  where it was shown that to get physical behavior one needs to include also a contribution due to orbital magnetization. Below, we predict a contribution to the spin Nernst conductivity due to a spin-resolved orbital magnetization, that restores physical behavior.

%==========================================================
\subsection{Contribution due to orbital spin-resolved magnetization}
%==========================================================

In the case of 2DEG with Rashba spin-orbit interaction there is no usual orbital magnetization, which is suppressed by time-reversal symmetry. Instead of this, we predict a spin-resolved orbital magnetization, $M\sigma_z$ ($M$ describes magnitude of the spin-resolved magnetization, referred to in the following as spin-resolved magnetization), which contributes to the spin current in thermal nonequilibrium, and thus also to the spin Nernst effect.\cite{Dyrdalpreprint}  Taking into account the
spin-resolved orbital magnetization, one should add a term $M(T)(\hbar/e)(\nabla T/T)$ to the spin current density, where $M(T)$ describes the spin-resolved  orbital magnetization at temperature $T$. The spin resolved orbital magnetization  and its temperature dependence is calculated in a separate paper.\cite{Dyrdalpreprint}
Thus, the total spin Nernst conductivity, $\alpha_{xy}^{s_{z},{\rm tot}}$, can be written as
\begin{equation}
\alpha_{xy}^{s_{z},{\rm tot}} = \alpha_{xy}^{s_{z},{\rm ren}} + \Delta\alpha_{xy}^{s_{z}},
\end{equation}
where $\alpha_{xy}^{s_{z},{\rm ren}}$ is the contribution derived in the bubble approximation plus vertex correction, while
\begin{equation}
\Delta\alpha_{xy}^{s_{z}}= \frac{1}{T}M(T)(\hbar/e)
\end{equation}
is the contribution from spin-resolved orbital magnetization.

In the limit of zero temperature and for $\mu_0 >0$, the spin-resolved orbital magnetization is
\begin{equation}
M(T\to 0)  = -\frac{em\alpha^2}{12\pi\hbar^3}
\end{equation}
Thus, the corresponding contribution to the spin Nernst conductivity  is
\begin{equation}
\Delta\alpha_{xy}^{s_{z}} = -\frac{m\alpha^2}{12\pi\hbar^2}\frac{1}{T},
\end{equation}
which cancels the first term in Eq.(69). Accordingly, $\alpha_{xy}^{s_{z},{\rm tot}}$ tends to zero for $T\to 0$,
\begin{equation}
\alpha_{xy}^{s_{z},{\rm tot}}(T=0)  = 0.
\end{equation}

\begin{figure}[t]
  \centering
  % \Requires \usepackage{graphicx}
  \includegraphics[width=0.97\columnwidth]{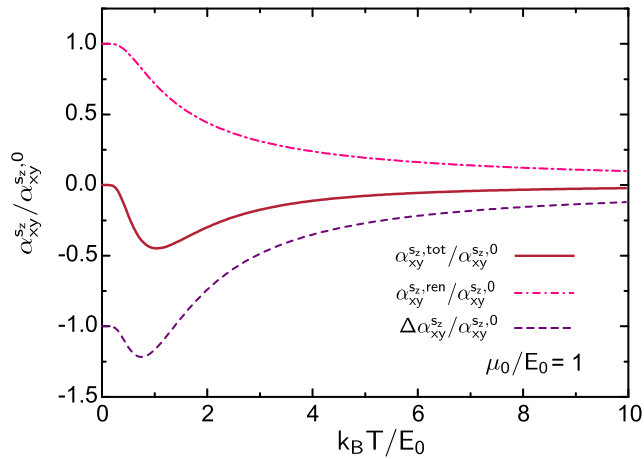}
  \caption{(Color online) Temperature dependence of the total spin Nernst conductivity, $\alpha_{xy}^{s_{z},{\rm tot}}$,  normalized to $\alpha_{xy}^{s_{z},0}$. The spin Nernst conductivity in the {\it bare bubble } approximation with the  vertex correction, $\alpha_{xy}^{s_{z},{\rm ren}}$, and the orbital contribution $\Delta\alpha_{xy}^{s_{z}}$ -- both normalized to $\alpha_{xy}^{s_{z},0}$ are also presented there. Other parameters as indicated and/or
  in Fig.2.}\label{SNEvertex}
\end{figure}

For a finite $T$, the spin resolved magnetization is given by the formula
\begin{eqnarray}
M(T) = \frac{\alpha e}{16 \pi \hbar} \left[ \int dk \left( \frac{2 \varepsilon_{k}^{2}}{\alpha^{2} k^{2}} - 1\right) [f(E_{+}) - f(E_{-})] \right.\nonumber\\
- \int dk \alpha k \left( \frac{2 \varepsilon_{k}^{2}}{\alpha^{2} k^{2}} - 1\right) [f'(E_{+}) + f'(E_{-})]\nonumber\\
+ \int dk \alpha^{2} k^{2} \left( \frac{2 \varepsilon_{k}^{2}}{\alpha^{2} k^{2}} - 1\right) [f''(E_{+}) - f''(E_{-})] \nonumber\\
+ \left. 2 \int dk \alpha k [E_{+} f''(E_{+}) + E_{-} f''(E_{-})] \right]\hspace{1.0cm}
\end{eqnarray}
From the above  formula and from Eq.(70) follows, that the total spin Nernst conductivity $\alpha_{xy}^{s_{z},{\rm tot}}$  is generally nonzero at finite $T$, but it vanishes when $T\to 0$.

In Fig.4 we show the
temperature dependence of the total spin Nernst conductivity, normalized to $\alpha_{xy}^{s_{z},0}$. We also show there the spin Nernst conductivity in the {\it bare bubble } approximation with the  vertex correction, $\alpha_{xy}^{s_{z},{\rm ren}}$, and the orbital contribution $\Delta\alpha_{xy}^{s_{z}}$ -- both normalized to $\alpha_{xy}^{s_{z},0}$. It is evident from this figure that $\alpha_{xy}^{s_{z},{\rm tot}}$ vanishes in the limit of $T=0$, and is generally nonzero for finite temperatures. The spin Nernst conductivity, however, decreases with increasing temperature and tends to zero for sufficiently high values of $T$.

%----------------------------------
\section{Conclusions}
%----------------------------------

In this paper we have calculated the temperature dependence of the spin Hall and spin Nernst conductivities
of a two-dimensional electron gas with Rashba spin-orbit interaction. To do this we have employed the approach based on the Matsubara Green functions.
The formalism used in the case of electric field as a driving force was subsequently adapted to the situation of spin current driven by a temperature gradient. To achieve this, we have  used the concept of an auxiliary  vector potential.

Both spin Hall and spin Nernst conductivities were calculated in the approximation including the vertex correction.  In the case of spin Nernst effect, we have also predicted a contribution to the spin Nernst conductivity following from a spin-resolved orbital magnetization. This term assures correct zero-temperature limit of the spin Nernst conductivity. When the vertex correction is included, the spin Hall conductivity vanishes exactly --  not only in the zero temperature limit, but also at finite temperatures, independently of the Rashba parameter, which is agreement with earlier predictions.\cite{Dimitrova2005,Erlingsson2005,Schliemann2006}  In turn, the spin Nernst conductivity vanishes at $T=0$, but remains generally nonzero at finite $T$.
It is worth noting, that our calculations concern constant (dc) spin Hall and spin Nernst conductivities, but the effects under consideration may be investigated also at nonzero frequencies.
\cite{Ma,Li2008,Iglesias2014}

\begin{acknowledgments}
This work was supported by the National Science Center in Poland as the Project
No. DEC-2012/04/A/ST3/00372. The authors would like to acknowledge very useful comments from Atsuo Shitade.
\end{acknowledgments}

\appendix

\begin{widetext}

\section{Integration over $\varepsilon$}

Taking into account Eqs~(22) and (23), one can write the
integrals over $\varepsilon$ that occur in Eqs (19) and (51), in the following form
\begin{eqnarray}
\int \frac{d \varepsilon}{2 \pi} f(\varepsilon) \mathcal{T}_{1}(\varepsilon + \omega, \varepsilon)
 = i \pi \alpha \frac{\hbar k}{2 m} \left[ \mathcal{I}^{RA}_{-+}(\varepsilon + \omega, \varepsilon) - \mathcal{I}^{RR}_{-+}(\varepsilon + \omega, \varepsilon)
  - \mathcal{I}^{RA}_{+-}(\varepsilon + \omega, \varepsilon) + \mathcal{I}^{RR}_{+-}(\varepsilon + \omega, \varepsilon) \right],
\end{eqnarray}
\begin{eqnarray}
\int \frac{d \varepsilon}{2 \pi} f(\varepsilon) \mathcal{T}_{2}(\varepsilon, \varepsilon - \omega)
= i \pi \alpha \frac{\hbar k}{2 m} \left[ \mathcal{I}^{AA}_{-+}(\varepsilon, \varepsilon - \omega) - \mathcal{I}^{AA}_{+-}(\varepsilon, \varepsilon - \omega)
 - \mathcal{I}^{RA}_{-+}(\varepsilon, \varepsilon - \omega) + \mathcal{I}^{RA}_{+-}(\varepsilon, \varepsilon - \omega) \right],
\end{eqnarray}
where $ \mathcal{I}^{\gamma \gamma^\prime}_{-+}$ and  $\mathcal{I}^{\gamma \gamma^\prime}_{+-}$ (for $\gamma =R,A$ and $\gamma^\prime  =R,A$) are integrals of products of retarded and/or advanced Green functions, as defined below. These integrals can be found using the identity $1/(x \pm i x') = \mathcal{P}{1/x} \mp i \pi \delta(x)$:
\begin{eqnarray}
\mathcal{I}_{-+}^{RA}(+\omega)= \int \frac{d \varepsilon}{2 \pi} f(\varepsilon)G_{-}^{R}(\varepsilon+\omega) G_{+}^{A}(\varepsilon)\hspace{9.5cm}\nonumber\\ = - \mathcal{P} \int \frac{d \varepsilon}{2 \pi} \frac{f(\varepsilon)}{E_{+} - E_{-} + \omega + 2 i \Gamma} \frac{1}{\varepsilon + \omega - E_{-} + \mu} + \mathcal{P} \int \frac{d \varepsilon}{2 \pi} \frac{f(\varepsilon)}{E_{+} - E_{-} + \omega + 2 i \Gamma} \frac{1}{\varepsilon + \mu - E_{+}}\nonumber\\
+ \frac{i}{2} \frac{f(E_{-} - \omega)}{E_{+} - E_{-} + \omega + 2 i \Gamma} + \frac{i}{2} \frac{f(E_{+})}{E_{+} - E_{-} + \omega + 2 i \Gamma}
\end{eqnarray}
\begin{eqnarray}
\mathcal{I}_{-+}^{RR}(+\omega) = \int \frac{d \varepsilon}{2 \pi} f(\varepsilon)G_{-}^{R}(\varepsilon+\omega) G_{+}^{R}(\varepsilon)\hspace{9.5cm}\nonumber\\ = - \mathcal{P} \int \frac{d \varepsilon}{2 \pi} \frac{f(\varepsilon)}{E_{+} - E_{-} + \omega} \frac{1}{\varepsilon + \omega + \mu - E_{-}} + \mathcal{P} \int \frac{d \varepsilon}{2 \pi} \frac{f(\varepsilon)}{E_{+} - E_{-} + \omega} \frac{1}{\varepsilon + \mu - E_{+}}\nonumber\\
+ \frac{i}{2} \frac{f(E_{-} - \omega)}{E_{+} - E_{-} + \omega} - \frac{i}{2} \frac{f(E_{+})}{E_{+} - E_{-} + \omega}
\end{eqnarray}
\begin{eqnarray}
\mathcal{I}_{+-}^{RA}(+\omega) = \int \frac{d \varepsilon}{2 \pi} f(\varepsilon)G_{+}^{R}(\varepsilon+\omega) G_{-}^{A}(\varepsilon)\hspace{9.5cm}\nonumber\\  = \mathcal{P} \int \frac{d \varepsilon}{2 \pi} \frac{f(\varepsilon)}{E_{+} - E_{-} - \omega - 2i\Gamma} \frac{1}{\varepsilon + \omega + \mu - E_{+}} - \mathcal{P} \int \frac{d \varepsilon}{2 \pi} \frac{f(\varepsilon)}{E_{+} - E_{-} - \omega - 2i\Gamma} \frac{1}{\varepsilon + \mu - E_{-}}\nonumber\\
- \frac{i}{2} \frac{f(E_{+} - \omega)}{E_{+} - E_{-} - \omega - 2 i \Gamma} - \frac{i}{2} \frac{f(E_{-})}{E_{+} - E_{-} - \omega - 2 i \Gamma}
\end{eqnarray}
\begin{eqnarray}
\mathcal{I}_{+-}^{RR}(+\omega) = \int \frac{d \varepsilon}{2 \pi} f(\varepsilon)G_{+}^{R}(\varepsilon+\omega) G_{-}^{R}(\varepsilon)\hspace{9.5cm}\nonumber\\  = \mathcal{P} \int \frac{d \varepsilon}{2 \pi} \frac{f(\varepsilon)}{E_{+} - E_{-} - \omega} \frac{1}{\varepsilon + \omega + \mu - E_{+}} - \mathcal{P} \int \frac{d \varepsilon}{2 \pi} \frac{f(\varepsilon)}{E_{+} - E_{-} - \omega} \frac{1}{\varepsilon + \mu - E_{-}}\nonumber\\
- \frac{i}{2} \frac{f(E_{+} - \omega)}{E_{+} - E_{-} - \omega} + \frac{i}{2} \frac{f(E_{-})}{E_{+} - E_{-} - \omega}
\end{eqnarray}
\begin{eqnarray}
\mathcal{I}_{-+}^{AA}( - \omega)= \int \frac{d \varepsilon}{2 \pi} f(\varepsilon)G_{-}^{A}(\varepsilon) G_{+}^{A}(\varepsilon - \omega)\hspace{9.5cm}\nonumber\\ = - \mathcal{P} \int \frac{d \varepsilon}{2 \pi} \frac{f(\varepsilon)}{E_{+} - E_{-} + \omega} \frac{1}{\varepsilon + \mu - E_{-}} + \mathcal{P} \int \frac{d \varepsilon}{2 \pi} \frac{f(\varepsilon)}{E_{+} - E_{-} + \omega} \frac{1}{\varepsilon + \mu - \omega - E_{+}} \nonumber\\
- \frac{i}{2} \frac{f(E_{-})}{E_{+} - E_{-} + \omega} + \frac{i}{2} \frac{f(E_{+} + \omega)}{E_{+} - E_{-} + \omega}
\end{eqnarray}
\begin{eqnarray}
\mathcal{I}_{+-}^{AA}( - \omega) = \int \frac{d \varepsilon}{2 \pi} f(\varepsilon)G_{+}^{A}(\varepsilon) G_{-}^{A}(\varepsilon - \omega)\hspace{9.5cm}\nonumber\\  = \mathcal{P} \int \frac{d \varepsilon}{2 \pi} \frac{f(\varepsilon)}{E_{+} - E_{-} - \omega} \frac{1}{\varepsilon + \mu - E_{+}} - \mathcal{P} \int \frac{d \varepsilon}{2 \pi} \frac{f(\varepsilon)}{E_{+} - E_{-} - \omega} \frac{1}{\varepsilon + \mu - \omega - E_{-}}\nonumber\\
\frac{i}{2} \frac{f(E_{+})}{E_{+} - E_{-} - \omega} - \frac{i}{2} \frac{f(E_{-} + \omega)}{E_{+} - E_{-} - \omega}
\end{eqnarray}
\begin{eqnarray}
\mathcal{I}_{-+}^{RA}(- \omega)  = \int \frac{d \varepsilon}{2 \pi} f(\varepsilon)G_{-}^{R}(\varepsilon) G_{+}^{A}(\varepsilon - \omega)\hspace{9.5cm} \nonumber\\ = - \mathcal{P} \int\frac{d \varepsilon}{2 \pi} \frac{f(\varepsilon)}{E_{+} - E_{-} + \omega + 2i \Gamma} \frac{1}{\varepsilon + \mu - E_{-}} + \mathcal{P} \int \frac{d \varepsilon}{2 \pi} \frac{f(\varepsilon)}{E_{+} - E_{-} + \omega + 2 i \Gamma} \frac{1}{\varepsilon + \mu - \omega - E_{+}} \nonumber\\
\frac{i}{2} \frac{f(E_{-})}{E_{+} - E_{-} + \omega + 2 i \Gamma} + \frac{i}{2} \frac{f(E_{+} + \omega)}{E_{+} - E_{-} + \omega + 2 i \Gamma}
\end{eqnarray}
\begin{eqnarray}
\mathcal{I}_{+-}^{RA}(- \omega) = \int \frac{d \varepsilon}{2 \pi} f(\varepsilon)G_{+}^{R}(\varepsilon) G_{-}^{A}(\varepsilon - \omega)\hspace{9.5cm} \nonumber\\ = \mathcal{P} \int \frac{d \varepsilon}{2 \pi} \frac{f(\varepsilon)}{E_{+} - E_{-} - \omega - 2i \Gamma} \frac{1}{\varepsilon + \mu - E_{+}} - \mathcal{P} \int \frac{d \varepsilon}{2 \pi} \frac{f(\varepsilon)}{E_{+} - E_{-} - \omega - 2i \Gamma} \frac{1}{\varepsilon + \mu - \omega -E_{-}}\nonumber\\
-\frac{i}{2} \frac{f(E_{+})}{E_{+} - E_{-} - \omega - 2 i \Gamma} - \frac{i}{2} \frac{f(E_{-} + \omega)}{E_{+} - E_{-} - \omega - 2i \Gamma}
\end{eqnarray}

In turn, inserting Eqs~(A1) and (A2) into Eqs (19) and (51) one can rewrite the formula for spin Hall conductivity in the form
\begin{eqnarray}
\sigma_{xy}^{s_{z}} = - i \pi \alpha \frac{e \hbar}{\omega} \int \frac{d k}{(2 \pi)^{2}} \frac{\hbar k^{2}}{2 m} \mathcal{S}
\end{eqnarray}
and the formula for the spin Nernst conductivity as
\begin{equation}
\alpha_{xy}^{s_{z}} = - i \pi \alpha \frac{\hbar}{\omega} \frac{1}{T} \int \frac{dk}{(2\pi)^{2}} \frac{\hbar k^{2}}{2 m} (\varepsilon_{k} - \mu) \mathcal{S},
\end{equation}
where $\mathcal{S}$ is defined as follows:
\begin{eqnarray}
\mathcal{S} = \mathcal{I}^{RA}_{-+}(+ \omega) - \mathcal{I}^{RR}_{-+}(+ \omega)
- \mathcal{I}^{RA}_{+-}(+ \omega) + \mathcal{I}^{RR}_{+-}(+ \omega)
\nonumber\\
+\mathcal{I}^{AA}_{-+}( - \omega) - \mathcal{I}^{AA}_{+-}( - \omega)
 - \mathcal{I}^{RA}_{-+}( - \omega) + \mathcal{I}^{RA}_{+-}( - \omega).
\end{eqnarray}
Taking now into account Eqs (A3) to (A10) one finds
\begin{eqnarray}
\mathcal{S} = i\omega \left[ - \frac{(E_{+} - E_{-})[f'(E_{+}) + f'(E_{-})]}{(E_{+} - E_{-})^{2} - (\omega + 2 i \Gamma)^{2}} + \frac{(E_{+} - E_{-}) [f'(E_{+}) + f'(E_{-})]}{(E_{+} - E_{-})^{2} + \omega^{2}} - 2 \frac{f(E_{+}) - f(E_{-})}{(E_{+} - E_{-})^{2} - \omega^{2}}\right]
\end{eqnarray}
Inserting now Eq.(A14) into Eqs (A11) and (A12)], and taking the limit $\omega \rightarrow 0$,  one arrives at Eq.(24) and  Eq.(52) for the spin Hall and spin Nernst conductivities, respectively.

\end{widetext}

%=====================================================
\section{Self-energy}
%=====================================================
In this appendix we calculate the self-energy and relaxation time due to scattering on point-like impurities
in the Born approximation. The self-energy (at the Fermi level) is given by the following equation:
\begin{equation}
\Sigma_{\mathbf{k}}^{R} = n_{i}V^{2} \int \frac{d^{2} \mathbf{k}}{(2\pi)^{2}} G^{R}_{\mathbf{k}},
\end{equation}
where $n_i$ is the impurity concentration, $V$ is the scattering potential od the impurity, and $G^{R}_{\mathbf{k}}$ is the Green function taken at the Fermi level.
Writing $k_x=k\cos \phi$ and $k_y=k\sin \phi$, and integration over $\phi$ one finds the formula
\begin{equation}
\Sigma_{\mathbf{k}}^{R} = \pi n_{i}V^{2} \sigma_{0} \int \frac{dk k}{(2\pi)^{2}} (G_{+}^{R} + G_{-}^{R}),
\end{equation}
from which one finds
\begin{eqnarray}
\Sigma_{\mathbf{k}}^{R} = \pi n_{i}V^{2} \sigma_{0} \int \frac{dk k}{(2\pi)^{2}} \left[ \mathcal{P}\left( \frac{1}{\mu - E_{+}}\right) - i \pi \delta(\mu - E_{+}) \right.\nonumber\\
+ \left. \mathcal{P}\left( \frac{1}{\mu - E_{-}}\right)   - i \pi \delta(\mu - E_{-}) \right],\hskip0.7cm
\end{eqnarray}
where $\mathcal{P}$ indicates principal value of the corresponding integral.
Since $\Sigma_{\mathbf{k}}^{R} = \Re(\Sigma_{\mathbf{k}}^{R}) - i \Gamma \sigma_{0}$, we find from the above formula
\begin{eqnarray}
\Gamma = n_{i}V^{2} \pi^{2} \int \frac{dk k}{(2\pi)^{2}} \left[ \delta(\mu - E_{+}) + \delta(\mu - E_{-})\right].
\end{eqnarray}

For  $\mu>0$, the integrals in the above equation can be easily calculated and are
\begin{eqnarray}
\int \frac{dk k}{(2\pi)^{2}} \delta(\mu - E_{\pm})
= \frac{m}{4 \pi^{2} \hbar^{2}} \left(1 \mp  \frac{m \alpha}{\sqrt{2 m \mu \hbar^{2} + m^{2} \alpha^{2}}}\right),\qquad
\end{eqnarray}
so the relaxation rate $\Gamma$ for $\mu>0$ is
\begin{equation}
\Gamma = \pi n_{i} V^{2} \frac{m}{2 \pi \hbar^{2}}.
\end{equation}
Note, $\Gamma$ is here independent of the chemical potential.

In turn, for  $\mu < 0$ one finds
\begin{equation}
\int \frac{dk k}{(2\pi)^{2}} \delta(\mu - E_{+}) = 0
\end{equation}
and
\begin{eqnarray}
\int \frac{dk k}{(2\pi)^{2}} \delta(\mu - E_{-})
= \frac{m (k_{-}^{+} + k_{-}^{-})}{\sqrt{2 m \mu \hbar^{2} + m^{2} \alpha^{2}}}.
\end{eqnarray}
Thus, $\Gamma$ is now given by the formula
\begin{equation}
\Gamma = \pi n_{i} V^{2} \frac{m}{2 \pi \hbar^{2}} \frac{n^{\ast}}{n},
\end{equation}
where $n$ and $n^*$ are electron concentrations defined in Sec.III below Eq.(31). Note,
for $\mu<0$, $\Gamma$ depends on the chemical potential.

\begin{widetext}
%=====================================================
\section{Calculation of the vertex correction}
%=====================================================
Consider the second term on the right-hand side of Eq.(39). Inserting Eq.(38) into this term  and integrating over the angle $\phi$, this term takes the form
\begin{eqnarray}
\int d\phi\, G_{\mathbf{k}}^{A}(\varepsilon) \hat{J}_{x}^{s_{z}}G_{\mathbf{k}}^{R}(\varepsilon + \omega) =
\frac{\pi}{2} \left\{\left(G_{+}^{A}(\varepsilon) + G_{-}^{A}(\varepsilon)\right) \left(G_{+}^{R}(\varepsilon + \omega) + G_{-}^{R}(\varepsilon + \omega) \right) b
\right.\hspace{5cm}\nonumber\\ \left.
+ i \frac{k}{2} \left[\, \left(G_{+}^{A}(\varepsilon) + G_{-}^{A}(\varepsilon)\right) \left(G_{+}^{R}(\varepsilon + \omega) - G_{-}^{R}(\varepsilon + \omega)\right)
- \left(G_{+}^{A}(\varepsilon) - G_{-}^{A}(\varepsilon)\right) \left(G_{+}^{R}(\varepsilon + \omega) + G_{-}^{R}(\varepsilon + \omega)\right) \, \right] d \right\} \sigma_{x}\hspace{0.3cm}\nonumber\\ \nonumber\\
 +\frac{\pi}{2} \left(G_{+}^{A}(\varepsilon) + G_{-}^{A}(\varepsilon)\right) \left(G_{+}^{R}(\varepsilon + \omega) + G_{-}^{R}(\varepsilon + \omega)\right) c \, \sigma_{y} \hspace{0.3cm}\nonumber\\ \nonumber\\
 +\frac{\pi}{2}\left[\, \left(G_{+}^{A}(\varepsilon) - G_{-}^{A}(\varepsilon)\right) \left(G_{+}^{R}(\varepsilon + \omega) - G_{-}^{R}(\varepsilon + \omega)\right)
+  \left(G_{+}^{A}(\varepsilon) + G_{-}^{A}(\varepsilon)\right) \left(G_{+}^{R}(\varepsilon + \omega) + G_{-}^{R}(\varepsilon + \omega)\right) \, \right] a \, \sigma_{0}.\hspace{0.3cm}\nonumber\\
\end{eqnarray}
Let us introduce now the following notation:
\begin{eqnarray}
\int d\phi G_{\mathbf{k}}^{A}(\varepsilon) \hat{J}_{x}^{s_{z}} G_{\mathbf{k}}^{R}(\varepsilon + \omega) =
\frac{\pi}{2} \left( F_{x1}\, b\, + i \frac{k}{2} F_{x2}\, d\, \right)\sigma_{x}+ \frac{\pi}{2} F_{y}\, c\, \sigma_{y} + \frac{\pi}{2} F_{0}\, a\, \sigma_{0},\hspace{0.5cm}
\end{eqnarray}
where:
\begin{eqnarray}
\label{Fx1}
F_{x1} = F_{y}=\left(G_{+}^{A}(\varepsilon) + G_{-}^{A}(\varepsilon)\right)\left(G_{+}^{A}(\varepsilon + \omega) + G_{-}^{A}(\varepsilon + \omega)\right) \hspace{7cm}
\nonumber\\
= G_{+}^{A}(\varepsilon)G_{+}^{R}(\varepsilon + \omega) + G_{+}^{A}(\varepsilon)G_{-}^{R}(\varepsilon + \omega)
+ G_{-}^{A}(\varepsilon)G_{+}^{R}(\varepsilon + \omega) + G_{-}^{A}(\varepsilon)G_{-}^{R}(\varepsilon + \omega),
\end{eqnarray}
\begin{eqnarray}
\label{Fx2}
F_{x2} =  \left(G_{+}^{A}(\varepsilon) + G_{-}^{A}(\varepsilon)\right) \left(G_{+}^{R}(\varepsilon + \omega) - G_{-}^{R}(\varepsilon + \omega)\right)
-  \left(G_{+}^{A}(\varepsilon) - G_{-}^{A}(\varepsilon)\right)  \left(G_{+}^{R}(\varepsilon + \omega) + G_{-}^{R}(\varepsilon + \omega)\right)\nonumber\\
= 2 \left(G_{+}^{R}(\varepsilon + \omega)G_{-}^{A}(\varepsilon) - G_{-}^{R}(\varepsilon + \omega) G_{+}^{A}(\varepsilon)\right),
\end{eqnarray}
\begin{eqnarray}
F_{0} = \left(G_{+}^{A}(\varepsilon) - G_{-}^{A}(\varepsilon)\right) \left(G_{+}^{R}(\varepsilon + \omega) - G_{-}^{R}(\varepsilon + \omega)\right)
+ \left(G_{+}^{A}(\varepsilon) + G_{-}^{A}(\varepsilon)\right) \left(G_{+}^{R}(\varepsilon + \omega) + G_{-}^{R}(\varepsilon + \omega)\right) \nonumber\\
= 2 \left(G_{+}^{R}(\varepsilon + \omega)G_{+}^{A}(\varepsilon) + G_{-}^{R}(\varepsilon + \omega)G_{-}^{A}(\varepsilon) \right).
\end{eqnarray}

Taking into account Eq.(40), the equation for the spin current vertex (39) takes the form
\begin{eqnarray}
a\, \sigma_{0} + b\, \sigma_{x} + c\, \sigma_{y} + d k_{x} \sigma_{z} = a\, V_{0}^{2} \int \frac{dk k}{8 \pi} F_{0} \, \sigma_{0}
+ V_{0}^{2}  \int \frac{dk k}{(2\pi)^{2}} \left[ \frac{\pi}{2} F_{x1}\, b + i \frac{\pi}{4} k F_{x2}\, d\right]\, \sigma_{x}
\nonumber\\
+c\, V_{0}^{2}  \int \frac{dk k}{8\pi} F_{y}\, \sigma_{y} + \frac{\hbar^{2}}{2 m} k_{x}\, \sigma_{z}.
\end{eqnarray}
>From this equality we find
\begin{eqnarray}
a=c=0,\qquad
d = \frac{\hbar^{2}}{2 m},\qquad
\label{b}
b = \frac{i \frac{\hbar^{2} V_{0}^{2}}{8 m} \int \frac{dk k}{4 \pi}  k F_{x2}}{1 - V_{0}^{2} \int \frac{dk k}{8 \pi} F_{x1}} = \frac{i\, \mathcal{I}_{2}}{1 - \mathcal{I}_{1}},
\end{eqnarray}
with
\begin{eqnarray}
\mathcal{I}_{1} = n_{i}V^{2} \left( \mathcal{I}_{1++} + \mathcal{I}_{1-+} + \mathcal{I}_{1+-} + \mathcal{I}_{1--}\right),
\end{eqnarray}
\begin{equation}
\mathcal{I}_{2} = \frac{\hbar^{2} n_{i}V^{2}}{8 m} \left( \mathcal{I}_{2+-} -\, \mathcal{I}_{2-+}\right),
\end{equation}
where
\begin{eqnarray}
\mathcal{I}_{1 \pm \pm} = \int \frac{dk k}{8 \pi} G_{\pm}^{R}(\varepsilon+\omega) G_{\pm}^{A}(\varepsilon) =
\int \frac{dk k}{8 \pi} \frac{1}{\varepsilon + \omega + \mu - E_{\pm} + i \Gamma} \frac{1}{\varepsilon + \mu - E_{\pm} - i \Gamma},\hspace{1cm}\\
\mathcal{I}_{1+-} = \int \frac{dk k}{8 \pi} G_{+}^{R}(\varepsilon+\omega) G_{-}^{A}(\varepsilon)=
\int \frac{dk k}{8 \pi} \frac{1}{\varepsilon + \omega + \mu - E_{+} + i \Gamma} \frac{1}{\varepsilon + \mu - E_{-} - i \Gamma},\hspace{1cm}\\
\mathcal{I}_{1-+} = \int \frac{dk k}{8 \pi} G_{-}^{R}(\varepsilon+\omega) G_{+}^{A}(\varepsilon) =
\int \frac{dk k}{8 \pi} \frac{1}{\varepsilon + \omega + \mu - E_{-} + i \Gamma} \frac{1}{\varepsilon + \mu - E_{+} - i \Gamma} ,\hspace{1cm}\\
\mathcal{I}_{2+-} = \int \frac{dk k^{2}}{2\pi} G_{+}^{R}(\varepsilon+\omega) G_{-}^{A}(\varepsilon) =
\int \frac{dk k^{2}}{2\pi} \frac{1}{\varepsilon + \omega + \mu - E_{+} + i \Gamma} \frac{1}{\varepsilon + \mu - E_{-} - i \Gamma},\hspace{1cm}\\
\mathcal{I}_{2-+} = \int \frac{dk k^{2}}{2\pi} G_{-}^{R}(\varepsilon+\omega) G_{+}^{A}(\varepsilon) =
\int \frac{dk k^{2}}{2\pi} \frac{1}{\varepsilon + \omega + \mu - E_{-} + i \Gamma} \frac{1}{\varepsilon + \mu - E_{+} - i \Gamma}.\hspace{1cm}
\end{eqnarray}

Then, upon decomposing the expressions under integrals into simple fractions and putting $\varepsilon = 0$ and $\omega = 0$ we find
\begin{eqnarray}
\mathcal{I}_{1 \pm \pm} =
 \frac{\pi}{\Gamma} \left( \int \frac{dk k}{8 \pi} \delta(\mu  - E_{\pm}) + \int \frac{dk k}{8 \pi} \delta(\mu - E_{\pm})\right),
\end{eqnarray}
\begin{eqnarray}
\mathcal{I}_{1+-}
= \mathcal{P} \int \frac{dk k}{8 \pi} \frac{1}{2 \alpha k} \frac{1}{\mu  - E_{+}} - \mathcal{P} \int \frac{dk k}{8 \pi} \frac{1}{2 \alpha k} \frac{1}{\mu - E_{-}}
-\frac{i \pi}{2 \alpha} \left( \int \frac{dk}{8 \pi}\delta(\mu - E_{+}) + \int \frac{dk}{8 \pi}  \delta(\mu - E_{-})\right),
\end{eqnarray}
\begin{eqnarray}
\mathcal{I}_{1-+}
= \mathcal{P}\int \frac{dk k}{8 \pi} \frac{1}{2 \alpha k} \frac{1}{\mu - E_{+}} - \mathcal{P}\int \frac{dk k}{8 \pi} \frac{1}{2 \alpha k} \frac{1}{\mu - E_{-}}
+ \frac{i \pi}{2 \alpha} \left(\int \frac{dk}{8 \pi}  \delta(\mu - E_{-}) + \int \frac{dk}{8 \pi}  \delta(\mu - E_{+})  \right),\nonumber\\
\end{eqnarray}
\begin{eqnarray}
\mathcal{I}_{2+-} =
 \frac{1}{2\alpha}\mathcal{P}\int \frac{dk k}{2\pi} \frac{1}{\mu - E_{+}} - \frac{1}{2\alpha}\mathcal{P}\int \frac{dk k}{2\pi} \frac{1}{\mu - E_{-}}
- \frac{i \pi}{2 \alpha} \int \frac{dk k}{2\pi} \delta(\mu - E_{+})
-  \frac{i \pi}{2 \alpha} \int \frac{dk k}{2\pi} \delta(\mu - E_{-}),
\end{eqnarray}
\begin{eqnarray}
\mathcal{I}_{2-+} =
- \frac{1}{2\alpha} \mathcal{P} \int \frac{dk k}{2\pi} \frac{1}{\mu - E_{-}} + \frac{1}{2 \alpha } \mathcal{P} \int \frac{dk k}{2\pi} \frac{1}{\mu - E_{+}}
+ \frac{i \pi}{2 \alpha} \int \frac{dk k}{2\pi} \delta(\mu - E_{-}) + \frac{i \pi}{2 \alpha} \int \frac{dk k}{2\pi} \delta(\mu - E_{+}).
\end{eqnarray}
Thus, we get
\begin{eqnarray}
\mathcal{I}_{1} = n_{i}V^{2}\left[\frac{\pi}{\Gamma} \int \frac{dk k}{8 \pi} \left(\delta(E_{+} - \mu) +  \delta(E_{-} - \mu)\right)
+\pi \int \frac{dk k}{8 \pi} \frac{\Gamma}{\alpha^2 k^{2} + \Gamma^{2}} \left(\delta(E_{+} - \mu) +\delta(E_{-} - \mu)\right)\right],
\end{eqnarray}
\begin{eqnarray}
\mathcal{I}_{2} = - i \pi n_{i} V^{2} \frac{\hbar^{2}}{8 m} \int \frac{dk k}{2 \pi}\frac{\alpha k^{2}}{\alpha^{2} k^{2} +  \Gamma^{2}} \left[\delta(E_{+} - \mu) + \delta(E_{-} - \mu)\right].
\end{eqnarray}

In the case of $\mu > 0$ we have
\begin{eqnarray}
\delta(E_{\pm} - \mu) = \frac{m}{\sqrt{m^{2} \alpha^{2} + 2 m \mu \hbar}} \delta(k - k_{\pm}),
\end{eqnarray}
so Eq.(C20) takes form
\begin{eqnarray}
\mathcal{I}_{1} = \pi n_{i} V^{2}  \frac{1}{8 \pi} \frac{m}{\sqrt{m^{2} \alpha^{2} + 2 m \mu \hbar}} \left[ \frac{1}{\Gamma} (k_{+} + k_{-})
+\Gamma\left( \frac{k_{+}}{\alpha^{2} k_{+}^{2} + \Gamma^{2}} \frac{k_{-}}{\alpha^{2} k_{-}^{2} + \Gamma^{2}}\right) \right],
\end{eqnarray}
while $\mathcal{I}_{2}$ is given by
\begin{eqnarray}
\mathcal{I}_{2} = - i n_{i} V^{2} \frac{\hbar^{2}}{16 m} \frac{m \alpha}{\sqrt{m^{2} \alpha^{2} + 2 m \mu \hbar}}
\left[ \frac{k_{+}^{3}}{\alpha^{2} k_{+}^{2} + \Gamma^{2}}  +  \frac{k_{-}^{3}}{\alpha^{2} k_{-}^{2} + \Gamma^{2}}\right].
\end{eqnarray}
Finally we get
\begin{eqnarray}
b = \frac{\hbar^{2}}{2 m} \frac{\Gamma}{\alpha}.
\end{eqnarray}
Similar calculations for $\mu < 0$ give the same form of the parameter $b$.

\end{widetext}

%%%%%%%%%%%%%%%%%%%%%%%%%%%%%%%%%%%%%%%%%%%%%%%%%%%%%%%%%%%%%%%%%%%%%%%%%%%%%%%%%%%%%%%%%%%%%%%%%%%%%%%%%%


\begin{references}
%%%%%%%%%%%%%%%%%%%%%%%%%%%%%%%%%%%%%%%%%%%%%%%%%%%%%%%%%%%%%%%%%%%%%%%%%%%%%%%%%%%%%%%%%%%%%%%%%%%%%%%%%%
%
\bibitem{Dyakonov1971}
M. I. Dyakonov and V. I. Perel, Pis. Z. Eksp. Teor. Fiz. {\bf{13}}, 657 (1971)
[JETP Lett. {\bf{13}}, 467 (1971)].

\bibitem{Dyakonov1971_2}
M. I. Dyakonov and V. I. Perel, Phys. Lett. A {\bf 35}, 459 (1971).

\bibitem{Hirsch1999}
J. E. Hirsch, \prl {\bf{83}}, 1834 (1999).

\bibitem{EngelRashbaHalperin}
H. A. Engel, E. I. Rashba, and B. I. Halperin, {\it{Handbook of
Magnetism and Advanced Magnetic Materials}} in Spintronics and Magnetoelectronics, vol. 5,
ed. H. Kronmuller and S. Parkin (New York: Wiley 2007), pp. 2858--2877.

\bibitem{SinovaUniversalSHE}
J. Sinova, D. Culcer, Q. Niu, N. A. Sinitsyn, T. Jungwirth, and A. H. MacDonald, \prl {\bf 92}, 126603 (2004).

\bibitem{Murakami2003}
S. Murakami, N. Nagaosa, S.-C. Zhang, Science {\bf 301}, 1348 (2003).

\bibitem{Kato2004}
Y. K. Kato, R. C. Myers, A. C. Gossard, and D. D. Awschalom, Science {\bf 306}, 1910 (2004).

\bibitem{Wunderlich2005}
J. Wunderlich, B. Kaestner, J. Sinova, and T. Jungwirth,
\prl {\bf 94}, 047204 (2005).

\bibitem{sinova2012}
J. Sinova and I. Zutic, Nature Materials {\bf{11}}, 368 (2012).

\bibitem{sinova2014}
J. Sinova, S. O. Valenzuela , J. Wunderlich, C. H. Back, and T. Jungwirth, arXiv: 1411.3249v1 (2014).

\bibitem{LiuMoriyama2011}
L. Liu, T. Moriyama, D.C. Ralph, and R. A. Buhram, \prl {\bf 106}, 036601 (2011).

\bibitem{LiuLee2012}
L. Liu, O.J. Lee, T. J. Gudmundsen, D.C. Ralph, and R. A. Buhrman, \prl {\bf 107}, 096602 (2012).

\bibitem{Pai2012}
Ch.-F. Pai, L. Liu, Y. Li, H. W. Tseng, D.C. Ralph, and R. A. Buhrman, Appl. Phys. Lett. {\bf 101}, 122404 (2012).

\bibitem{Vignale2010}
G. Vignale , J. Supercond. Nov. Magn. {\bf{23}}, 3 (2010).

\bibitem{Ma}
Z. Ma, Solid State Communications {\bf 150}, 510 (2010).

\bibitem{LiuXie}
X. Liu and X.C. Xie,  Solid State Communications {\bf 150}, 471 (2010).

\bibitem{Cheng2008}
S. G. Cheng, Y. Xing, Q. F. Sun, and X. C. Xie, \prb {\bf 78}, 045302 (2008).

\bibitem{Vignale2008}
K. Bencheikh and G. Vignale, \prb {\bf 77}, 155315 (2008).

\bibitem{Dimitrova2005}
O. V. Dimitrova, \prb {\bf 71}, 245327 (2005).

\bibitem{Erlingsson2005}
S.I. Erlingsson, J. Schliemann, and D. Loss Phys. Rev. B {\bf 71}, 035319 (2005).

\bibitem{Schliemann2006}
J. Schliemann, Int. J. Mod. Phys. B 20, 1015 (2006)

\bibitem{Lyapilin}
I. I. Lyapilin, Low Temp. Phys. {\bf 39}, 957 (2013).

\bibitem{Akera}
H. Akera and H. Suzuura, \prb {\bf 87}, 075301 (2013).

\bibitem{Tauber2013}
K. Tauber, D. V. Fedorov, M. Gradhand, and I. Mertig, \prb {\bf 87}, 161114(R) (2013).

\bibitem{Wimmer2013}
S. Wimmer, D. K\"{o}dderitzsch, K. Chadova, and H. Ebert, \prb {\bf 88}, 201108(R) (2013).

\bibitem{Zimmermann2014}
B. Zimmermann, K. Chadova, D. K\"{o}dderitzsch, S. Blugel, H. Ebert, D. V. Fedorov, N. H. Long, P. Mavropoulos,
I. Mertig, Y. Mokrousov, and M. Gradhand, \prb {\bf 90}, 220403(R) (2014).

\bibitem{Kovacik2015}
R. Kov\'{a}\v{c}ik, P. Mavropoulos and S. Bl\"{u}gel, \prb {\bf 91}, 014421 (2015).

\bibitem{Tolle2014}
S. Tolle, C. Gorini, and U. Eckern, \prb {\bf{90}}, 235117 (2014).

\bibitem{Borge}
J. Borge, C. Gorini, and R. Raimondi, Phys. Rev. B {\bf 87}, 085309 (2013).


\bibitem{RotheHankiewicz}
D. G. Rothe, E. M. Hankiewicz, B. Trauzettel, and M. Guigou, \prb {\bf 86}, 165434 (2012).

\bibitem{Gorini2015}
C. Gorini, U. Eckern, and R. Raimondi,
\prl {\bf 115}, 076602 (2015).

\bibitem{obraztsov} Yu. N. Obraztsov, Fiz. Tverd. Tela {\bf 6}, 414 (1964) [Sov. Phys. Solid State {\bf 6}, 331 (1964)].


\bibitem{gusynin} V. P. Gusynin, S. G. Sharapov, and A. A. Varlamov, \prb {\bf{90}}, 155107 (2014).


\bibitem{Dyrdal2013}
A. Dyrda{\l}, M. Inglot, V. K. Dugaev, and J. Barna\'s, \prb {\bf{87}}, 245309 (2013).

\bibitem{mahan}
G. D. Mahan, \textit{Many Particle Physics}  (Kluwer Academic/Plenum Publishers, New York, 2000).

\bibitem{Luttinger}
J. M. Luttinger, Phys. Rev. {\bf 135}, A1505 (1964).

\bibitem{Strinati}
G. Strinati and C. Castellani, \prb {\bf 36}, 2270 (1987).

\bibitem{Tatara2015_1}
G. Tatara,
%Thermal vector potential theory of transport induced by a temperature gradient.
\prl {\bf 114}, 196601 (2015).

\bibitem{Tatara2015_2}
G. Tatara,
\prb {\bf 92}, 064405 (2015).

\bibitem{Borosco}
V. Brosco, L. Benfatto, E. Cappelluti, and C. Grimaldi, Phys. rev. Lett. {\bf 116}, 166602 (2016).


\bibitem{Inoue2004}
J. Inoue, G.E. Bauer, L.W. Molenkamp, \prb  {\bf 70}, 041303 (2004).

\bibitem{ChalaevLoss2004}
 O. Chalaev, D. Loss, \prb {\bf 71}, 245318 (2004).

\bibitem{Mishchenko2004}
E. Mishchenko, A. Shytov, B. Halperin, \prl {\bf 93}, 226602 (2004).

\bibitem{Tse}
W,-K. Tse and S. Das Sarma, Phys. Rev. B {\bf 74}, 245309 (2006).

\bibitem{Xiao2006}
Di Xiao, Y. Yao, Z. Fang, and Q. Niu, \prl {\bf 97}, 026603 (2006).

\bibitem{Chuu}
Ch.-P. Chuu, M-Ch. Chang, and Q. Niu, Solid State Communications {\bf 150}, 533 (2010).

\bibitem{Dyrdalpreprint}
A. Dyrda{\l}, J. Barna\'s, and V. K. Dugaev, In preparation

\bibitem{Li2008}
Z. Li, Z. Ma, and C. Zhang, EPL {\bf 82}, 67003 (2008).

\bibitem{Iglesias2014}
P. E. Iglesias and J. A. Maytorena, \prb {\bf 89}, 155432 (2014).



\end{references}
\end{document}